%% file: main.tex
\documentclass[acmsmall,screen,nonacm]{acmart}

\input{header.tex}

\begin{document}

\title{SecReason: Vulnerability Detection and In-Stream Self-Repair for LLM Code Generation}

\author{Ruidong Han}
\affiliation{%
  \institution{Singapore Management University}
  \country{Singapore}}
\email{rdhan@smu.edu.sg}
\orcid{0000-0001-6859-6005}

\author{Zhensu Sun}
\affiliation{%
  \institution{Singapore Management University}
  \country{Singapore}}
\email{zhensuuu@gmail.com}

\author{Chengyan Ma}
\affiliation{%
  \institution{Singapore Management University}
  \country{Singapore}}
\email{chengyanma@smu.edu.sg}

\author{Yuqing Niu}
\affiliation{%
  \institution{Singapore Management University}
  \country{Singapore}}
\email{yuqingniu@smu.edu.sg}

\author{Zhou Yang}
\affiliation{%
  \institution{University of Alberta}
  \city{Edmonton}
  \country{Canada}}
\email{zy25@ualberta.ca}

\author{Siqi Ma}
\affiliation{%
  \institution{University of Wollongong}
  \city{Wollongong}
  \country{Australia}}
\email{siqim@uow.edu.au}
\orcid{0000-0003-3479-5713}

\author{David Lo}
\affiliation{%
  \institution{Singapore Management University}
  \country{Singapore}}
\email{davidlo@smu.edu.sg}
\orcid{0000-0002-4367-7201}

\renewcommand{\shortauthors}{Han et al.}

\begin{abstract}
\input{tex/abstract.tex}
\end{abstract}

\ccsdesc[500]{Software and its engineering~Software security engineering}
\ccsdesc[500]{Software and its engineering~Software testing and debugging}
\ccsdesc[300]{Computing methodologies~Natural language generation}

\keywords{large language models, code generation, vulnerability detection, static analysis, self-repair}

\maketitle

\section{Introduction}
\label{sec:intro}
\input{tex/introduction.tex}

\input{tex/motivation.tex}

\section{Methodology of \system}
\label{sec:system}
\input{tex/design.tex}

\section{Evaluation}
\label{sec:evaluation}
\input{tex/evalution.tex}

\section{Threats to Validity}
\label{sec:validity}
\input{tex/validity.tex}

\section{Related Work}
\label{sec:related}
\input{tex/related.tex}

\section{Conclusion}
\label{sec:conclusion}
\input{tex/conclusion.tex}

\bibliographystyle{ACM-Reference-Format}
\bibliography{ref}

\end{document}

%% file: header.tex
\usepackage{tabularx}
\usepackage{amsmath,amsfonts}
\usepackage{enumitem}
\usepackage{xspace}
\usepackage{needspace}
\usepackage{caption}

\usepackage{colortbl}
\usepackage{graphicx}
\usepackage{tikz}
\usetikzlibrary{positioning,calc,arrows.meta}
\usepackage{pifont}

\usepackage{booktabs}
\usepackage{multirow}
\usepackage{threeparttable}
\usepackage{makecell}
\usepackage{siunitx}

\usepackage[most,listings]{tcolorbox}
\usepackage{algorithm}
\usepackage{algpseudocode}

\usepackage{float}
\usepackage{subfig}

\DeclareCaptionType{code}[Listing]

\newcommand{\system}{\textsc{SecReason}\xspace}

\newsavebox{\myfuncbox}
\newcommand{\func}[1]{%
    \begin{lrbox}{\myfuncbox}%
    \lstinline[basicstyle=\ttfamily\fontsize{8.5pt}{10pt}\selectfont, columns=flexible, basewidth=0.45em, literate={-}{-}1 {"}{{\texttt{"}}}1]{#1}%
    \end{lrbox}%
    \scalebox{0.85}{\usebox{\myfuncbox}}%
}
\newcommand{\dquote}[1]{\textquotedblleft#1\textquotedblright}
\newcommand{\tool}[1]{\lstinline[basicstyle=\ttfamily\fontsize{8pt}{10pt}\selectfont, literate={-}{-}1]!#1!}

\tcbset{
	mylistingbox/.style={
		enhanced,
		arc=1pt,
		breakable=false,
		colback=white,
		colframe=cyan!45!gray,
		colbacktitle=cyan!15!white,
		coltitle=cyan!60!black,
		boxsep=2pt,
		boxrule=0.6pt,
		titlerule=0.6pt,
		top=1pt,
		bottom=1pt,
		left=1pt,
		right=1pt,
		fonttitle=\bfseries\small,
		halign title=center,
		drop fuzzy shadow=black!15,
	}
}

%% file: tex/abstract.tex
Large language models (LLMs) may introduce security vulnerabilities into generated code, which can persist into deployed systems when verification focuses on functional correctness.
Security prompts and retrieved security knowledge act before any code exists, constrained decoding limits token choices but cannot guide repair, and post-hoc repair must regenerate the full response.
None of them inspects the code during generation.
Reasoning LLMs draft code early in their chain of thought, well before delivery.
We present \system, an in-stream defense that monitors this pre-delivery window and inspects the code during generation.
Dual-track incremental scanning applies security rules, checking each completed statement as it is emitted and tracing dataflows across the accumulated code.
When a security rule fires, \system interrupts generation, inserts an in-stream warning, and resumes decoding, so the model repairs the draft within the same reasoning trace.
On tasks where unassisted generation introduces vulnerabilities, \system raises the share of runs delivering secure and functionally correct code from below 15\% to 72.8\%--98.1\% on \tool{CWEval} and \tool{SecCodeBench} under a corpus-calibrated ruleset.
\system consumes fewer tokens than post-hoc repair and, when security rules misfire on secure code, shows smaller functional regressions.

%% file: tex/introduction.tex
Developers increasingly delegate implementation tasks to code assistants and coding agents built on large language models (LLMs), but the generated code can introduce security vulnerabilities such as injection and weak cryptography~\cite{pearce2022asleep,fu2025security,ramirez2025state}.
Because developers primarily verify functionality and rarely screen for security, they tend to trust code that executes correctly~\cite{perry2023assistants,khalid2026trust}, so functionally correct but vulnerable code persists in software repositories.
In automated environments, agents write and execute code without intermediate human inspection~\cite{yang2024sweagent,wang2024codeact,wang2025openhands,zhang2024autocoderover}, so such code reaches execution undetected.

\input{fig_source/model-stream.tex}

Existing defenses act before, during, or after generation, but none inspects the code during generation.
Security fine-tuning~\cite{he2024safecoder}, security prompts~\cite{sajadi2025security}, and retrieved security knowledge~\cite{lin2025securitycourse} act before any code exists, so they cannot verify whether the delivered code is secure or name the flaw in a specific draft.
Constrained decoding methods act during generation, enforcing grammar~\cite{ugare2024syncode}, conditioning on execution feedback~\cite{lavon2025egcfg}, or blocking completions matching known weakness patterns~\cite{storhaug2023efficient}, but none can guide the model to repair its code.
Post-generation defenses, such as static scanners~\cite{sun2024gptscan,li2025iris} and automated program repair~\cite{pearce2023repair,fdsp2025,yang2026seecurepair}, inspect code precisely but only after generation completes.
Static analyzers raise many false alarms~\cite{johnson2013why,sadowski2018lessons}, so a checker that inspects every generation often flags code that is in fact secure.
Repairing a flagged vulnerability then requires regenerating the full response, which increases token consumption and risks breaking working code, and this cost recurs even when the alarm was false.

Reasoning LLMs~\cite{openai2024o1,gemini3,deepseekr1} make the code inspectable during generation.
Before producing a final answer, these models emit an extended chain of thought~\cite{wei2022cot}, and the implementation is typically drafted inside this trace, not after it (Figure~\ref{fig:model-stream}, left).
We measure how often the models deliver insecure code, how much of the delivered code the reasoning drafts already contain, and how early these drafts finish.
The study covers six open-weight models, which expose complete reasoning streams, and 173 tasks from two security benchmarks, \tool{CWEval}~\cite{peng2025cweval} and \tool{SecCodeBench}~\cite{chen2026seccodebench}.
Section~\ref{app:motivating} narrates the example behind the figure, and Section~\ref{app:empirical} details the protocol and reports the per-model results.

Across the six models, 13.2\% to 35.6\% of functionally valid generations on \tool{CWEval} and 42.6\% to 61.3\% on \tool{SecCodeBench} fail security tests.
Security terms do appear in these traces, but among runs whose reasoning mentions them, 5.4\% to 61.1\% still deliver insecure code.

In these same traces, 52.4\% to 99.8\% of delivered lines appear verbatim in a fenced code block of the reasoning, and five of the six models exceed 73\% on both benchmarks.
The best-matching draft closes at 45.2\% to 89.1\% of the trace, leaving 203 to 6,028 reasoning tokens on average before delivery.
The interval between drafting and delivery is therefore a pre-delivery window in which the code that will be delivered already exists.

We build on this observation and introduce \emph{in-stream detection}, which runs the checker during generation, not after it.
The checker scans the draft as it is generated, statement by statement.
When it reports a potential vulnerability, a short warning is inserted into the ongoing reasoning.
The model then decides how to respond, fixing the draft before the final code if the alarm is real and verifying the flagged code if it is false, so no extra turn is needed.
Handling an alarm costs a few reasoning tokens, so false alarms become cheap, not costly.

Applying static analysis inside the reasoning stream is not straightforward.
Static analyzers expect complete, syntactically valid files, whereas reasoning streams interleave partial code with natural-language text.
An in-stream defense must recognize statements as they complete and translate structured analyzer findings into feedback that the model can act on within the ongoing trace.

To address these challenges (Section~\ref{app:challenges}), we present \system, an in-stream defense system that intercepts vulnerable code drafts during reasoning and guides the model to repair them before delivery (Figure~\ref{fig:model-stream}, right).
Operating at the inference serving layer without modifying model weights, \system parses the streaming reasoning trace incrementally to recognize completed statements.
To balance latency against detection depth, it scans a security ruleset on two tracks.
A statement-local track checks each completed statement as it is emitted, and a context-dependent track catches dataflows spanning the accumulated code.

We evaluate \system on the same 173 tasks in five programming languages. Streaming detection replays the generations of all six models, and in-stream self-repair runs on two locally deployed models over the tasks where unassisted generation introduces vulnerabilities.
On these tasks, in-stream self-repair matches or exceeds post-hoc repair in every benchmark and model combination, raising the share of runs delivering secure and functionally correct code from below $15\%$ to $72.8\%$--$98.1\%$ under a corpus-calibrated ruleset while consuming fewer tokens.
When security rules misfire on secure code, \system also shows smaller functional regressions than post-hoc repair.

This paper makes the following contributions.
\begin{itemize}[leftmargin=*]

\item An empirical study of six reasoning models on 173 tasks, showing that implementations are drafted early in the reasoning stream, while security-term mentions do not reliably prevent insecure delivery.

\item \system, an in-stream defense at the inference serving layer that recognizes completed statements in the streaming reasoning trace and inserts warnings prompting the model to repair vulnerable drafts before delivery.

\item An evaluation on \tool{CWEval} and \tool{SecCodeBench} showing that in-stream self-repair matches or exceeds post-hoc repair while consuming fewer tokens.

\end{itemize}

%% file: fig_source/model-stream.tex
\begin{figure*}[htb]
\centering
\resizebox{\textwidth}{!}{%
\begin{tikzpicture}[
    x=1pt, y=-1pt,
    font=\sffamily,
    panel/.style={draw=panelborder, fill=panelbg, rounded corners=3pt, line width=0.6pt},
    promptpanel/.style={draw=promptborder, fill=promptbg, rounded corners=3pt, line width=0.6pt},
    delivpanel/.style={draw=promptborder, fill=delivbg, line width=0.6pt},
    fence/.style={draw=fencepurple, densely dashed, rounded corners=3pt, line width=0.6pt},
    codeline/.style={anchor=west, inner sep=0pt, font=\ttfamily\fontsize{7.3pt}{8.6pt}\selectfont},
    bulletline/.style={anchor=west, inner sep=0pt, font=\sffamily\fontsize{7.2pt}{8.4pt}\selectfont},
    sectlabel/.style={anchor=west, inner sep=0pt, font=\sffamily\fontsize{7.2pt}{8.4pt}\selectfont\bfseries, text=black!55},
    pill/.style={rounded corners=4.5pt, inner xsep=4.5pt, inner ysep=2.2pt, font=\sffamily\fontsize{6.6pt}{7.6pt}\selectfont\bfseries, text=white, align=center},
    tag/.style={anchor=center, inner sep=1.5pt, font=\sffamily\fontsize{6.8pt}{7.8pt}\selectfont\bfseries\itshape, text=fencepurple, fill=panelbg},
    warnlabel/.style={anchor=west, inner sep=1.5pt, font=\sffamily\fontsize{6.8pt}{7.8pt}\selectfont\bfseries, text=warnred, fill=panelbg},
    pauselabel/.style={anchor=center, inner sep=1.5pt, font=\sffamily\fontsize{6.4pt}{7.4pt}\selectfont\itshape, text=warnred, fill=panelbg},
    notetext/.style={anchor=west, inner sep=0pt, font=\sffamily\fontsize{7pt}{8.4pt}\selectfont\itshape, text=notered}
]
\definecolor{kwred}{RGB}{215,58,73}
\definecolor{strblue}{RGB}{10,48,105}
\definecolor{fnpurple}{RGB}{111,66,193}
\definecolor{fencepurple}{RGB}{150,115,166}
\definecolor{taggray}{RGB}{136,136,136}
\definecolor{omitgray}{RGB}{170,170,170}
\definecolor{darktext}{RGB}{51,51,51}
\definecolor{panelbg}{RGB}{247,247,252}
\definecolor{panelborder}{RGB}{213,213,228}
\definecolor{promptbg}{RGB}{249,249,251}
\definecolor{promptborder}{RGB}{204,204,204}
\definecolor{delivbg}{RGB}{251,252,251}
\definecolor{warnred}{RGB}{214,69,65}
\definecolor{notered}{RGB}{176,48,48}
\definecolor{secgreen}{RGB}{46,125,50}
\definecolor{hlred}{RGB}{248,211,211}
\definecolor{hlgreen}{RGB}{213,234,213}
\definecolor{barred}{RGB}{211,47,47}
\definecolor{bargreen}{RGB}{46,125,50}
\definecolor{licblue}{RGB}{108,142,191}
\definecolor{hdrgray}{RGB}{119,119,119}

\draw[-{Stealth[length=5pt]}, line width=0.8pt, black!70] (8,11) -- (8,260);
\node[rotate=90, anchor=center, inner sep=0pt, font=\sffamily\fontsize{6.8pt}{7.8pt}\selectfont\itshape, text=black!60] at (2.5,135) {Generation time};

\node[sectlabel] at (14,5) {PROMPT TASK};
\draw[promptpanel] (14,11) rectangle (438,32);
\node[codeline] at (18,17.5) {\textcolor{kwred}{def} \textcolor{fnpurple}{get\_url\_to\_visit}(domain: str, subdomain\_prefix: str) -> str:};
\node[codeline] at (18,25.5) {\hspace*{14.6pt}\textcolor{strblue}{\textquotedbl\textquotedbl\textquotedbl If the subdomain is empty or invalid, return the domain itself.\textquotedbl\textquotedbl\textquotedbl}};

\node[pill, fill=hdrgray, anchor=south west] at (14,48) {Unassisted generation};
\node[pill, fill=licblue, anchor=south west] at (230,48) {With SecReason};

\draw[panel] (14,53) rectangle (222,154);
\draw[panel] (230,53) rectangle (438,212);
\node[sectlabel, fill=panelbg, inner xsep=3.5pt, inner ysep=0.5pt] at (20,53) {REASONING TRACE};
\node[sectlabel, fill=panelbg, inner xsep=3.5pt, inner ysep=0.5pt] at (236,53) {REASONING TRACE};

\node[bulletline, text=omitgray] at (18,61) {* \ldots\ earlier reasoning omitted \ldots};
\node[bulletline, text=darktext] at (18,69) {* invalid = empty? \textquotesingle/\textquotesingle? \textquotesingle\#\textquotesingle? \textquotesingle?\textquotesingle? whitespace? dots? \ldots};
\node[bulletline, text=darktext] at (18,77) {* I\textquotesingle ll just check empty and \textquotesingle/\textquotesingle:};
\node[bulletline, text=omitgray] at (234,61) {* \ldots\ earlier reasoning omitted \ldots};
\node[bulletline, text=darktext] at (234,69) {* invalid = empty? \textquotesingle/\textquotesingle? \textquotesingle\#\textquotesingle? \textquotesingle?\textquotesingle? whitespace? dots? \ldots};
\node[bulletline, text=darktext] at (234,77) {* I\textquotesingle ll just check empty and \textquotesingle/\textquotesingle:};

\draw[fence] (18,83) rectangle (218,113);
\node[tag] at (198,83) {Draft};
\draw[fence] (234,83) rectangle (434,113);
\node[tag] at (374,83) {Draft};
\node[pill, fill=warnred, anchor=center, inner xsep=3.2pt, inner ysep=1.5pt, font=\sffamily\fontsize{6.2pt}{7pt}\selectfont\bfseries] at (415,83) {finding};
\node[codeline] at (22,89.5) {\textcolor{kwred}{if not} subdomain\_prefix \textcolor{kwred}{or} \textcolor{strblue}{\textquotesingle/\textquotesingle} \textcolor{kwred}{in} subdomain\_prefix:};
\fill[hlred] (234,85) rectangle (430,94);
\node[codeline] at (238,89.5) {\textcolor{kwred}{if not} subdomain\_prefix \textcolor{kwred}{or} \textcolor{strblue}{\textquotesingle/\textquotesingle} \textcolor{kwred}{in} subdomain\_prefix:};
\node[codeline] at (22,97.5) {\hspace*{14.6pt}\textcolor{kwred}{return} \textcolor{strblue}{f\textquotedbl{}https://\{domain\}\textquotedbl}};
\node[codeline] at (238,97.5) {\hspace*{14.6pt}\textcolor{kwred}{return} \textcolor{strblue}{f\textquotedbl{}https://\{domain\}\textquotedbl}};
\node[codeline] at (22,105.5) {\textcolor{kwred}{return} \textcolor{strblue}{f\textquotedbl{}https://\{subdomain\_prefix\}.\{domain\}\textquotedbl}};
\node[codeline] at (238,105.5) {\textcolor{kwred}{return} \textcolor{strblue}{f\textquotedbl{}https://\{subdomain\_prefix\}.\{domain\}\textquotedbl}};

\node[bulletline, text=darktext] at (18,120.5) {* This covers empty and path-like injection.};
\node[bulletline, text=omitgray] at (18,128.5) {* \ldots\ the reasoning continues \ldots};
\fill[barred] (14,134.5) rectangle (16.5,150.5);
\node[notetext] at (19,139) {The \textquotesingle\#\textquotesingle\ and \textquotesingle?\textquotesingle\ cases identified during planning are never};
\node[notetext] at (19,147) {revisited. The draft is delivered unchanged.};

\draw[draw=warnred, line width=0.7pt, densely dashed, rounded corners=3pt] (234,119) rectangle (434,145);
\node[warnlabel] at (238,119) {SecReason\,\textperiodcentered\ in-stream warning};
\node[pauselabel] at (398,119) {generation paused};
\node[anchor=west, inner sep=0pt, font=\ttfamily\fontsize{7pt}{8pt}\selectfont\bfseries, text=warnred] at (238,127) {[Security] CWE-918 SSRF};
\node[bulletline, text=darktext] at (238,136.5) {subdomain\_prefix flows into the URL unchecked.};
\draw[draw=warnred, line width=0.8pt] (425,131) ellipse (3.6 and 2.5);
\fill[warnred] (425,131) circle (1.1);
\draw[-{Stealth[length=3.5pt]}, draw=warnred, line width=0.8pt] (421,131) -- (419,131);

\node[anchor=west, inner sep=0pt, font=\sffamily\fontsize{6.4pt}{7.4pt}\selectfont\itshape, text=secgreen] at (234,151) {generation resumes};
\node[bulletline, text=darktext] at (234,159) {* Right, \textquotesingle\#\textquotesingle\ and \textquotesingle?\textquotesingle\ would terminate the host.};
\node[bulletline, text=darktext] at (234,167) {* Allow only letters, digits, and \textquotesingle-\textquotesingle.};

\node[tag] at (395,172) {Repaired draft};
\draw[fence] (234,172) rectangle (434,209);
\node[codeline] at (238,179) {\textcolor{kwred}{if not} subdomain\_prefix \textcolor{kwred}{or} \textcolor{kwred}{any}(};
\node[codeline] at (238,187) {\hspace*{14.6pt}c \textcolor{kwred}{in} \textcolor{strblue}{\textquotedbl/\#?\textquotedbl} \textcolor{kwred}{for} c \textcolor{kwred}{in} subdomain\_prefix):};
\node[codeline] at (238,195) {\hspace*{14.6pt}\textcolor{kwred}{return} \textcolor{strblue}{f\textquotedbl{}https://\{domain\}\textquotedbl}};
\node[codeline] at (238,203) {\textcolor{kwred}{return} \textcolor{strblue}{f\textquotedbl{}https://\{subdomain\_prefix\}.\{domain\}\textquotedbl}};

\draw[delivpanel] (14,163) rectangle (222,197);
\fill[barred] (14,163) rectangle (17,197);
\node[sectlabel, text=secgreen, fill=white, inner xsep=3.5pt, inner ysep=0.5pt] at (20,163) {DELIVERED CODE};
\node[pill, fill=warnred, anchor=center] at (198,163) {CWE-918};
\fill[hlred] (19,166) rectangle (218,175);
\node[codeline] at (22,170.5) {\textcolor{kwred}{if not} subdomain\_prefix \textcolor{kwred}{or} \textcolor{strblue}{\textquotesingle/\textquotesingle} \textcolor{kwred}{in} subdomain\_prefix:};
\node[codeline] at (22,178.5) {\hspace*{14.6pt}\textcolor{kwred}{return} \textcolor{strblue}{f\textquotedbl{}https://\{domain\}\textquotedbl}};
\node[codeline] at (22,186.5) {\textcolor{kwred}{return} \textcolor{strblue}{f\textquotedbl{}https://\{subdomain\_prefix\}.\{domain\}\textquotedbl}};

\draw[draw=warnred!60, fill=hlred!15, rounded corners=3pt, line width=0.6pt] (14,206) rectangle (222,259);
\node[sectlabel, text=warnred, fill=white, inner xsep=3.5pt, inner ysep=0.5pt] at (20,206) {EXPLOIT IMPACT};
\node[pill, fill=warnred, anchor=center, inner xsep=3pt, inner ysep=1.5pt, font=\sffamily\fontsize{6.2pt}{7pt}\selectfont\bfseries] at (198,206) {Bypass};
\node[codeline] at (18,214.5) {\textcolor{darktext}{Payload: }\textcolor{strblue}{\textquotedbl evil.com\#\textquotedbl}\textcolor{darktext}{ (passes check)}};
\node[codeline] at (18,223.5) {\textcolor{darktext}{URL: }\textcolor{strblue}{f\textquotedbl{}https://evil.com\#.\{domain\}\textquotedbl}};
\node[bulletline, text=warnred] at (18,233.5) {* \textquotesingle\#\textquotesingle\ terminates host in URL parsing:};
\node[bulletline, text=darktext] at (18,242.5) {\hspace*{4pt}Request fetches \textcolor{warnred}{\textbf{evil.com}} (CWE-918 SSRF).};

\draw[delivpanel] (230,221) rectangle (438,259);
\fill[bargreen] (230,221) rectangle (233,259);
\node[sectlabel, text=secgreen, fill=white, inner xsep=3.5pt, inner ysep=0.5pt] at (236,221) {DELIVERED CODE};
\node[pill, fill=secgreen, anchor=center] at (410,221) {secure};
\fill[hlgreen] (235,224.5) rectangle (434,241.5);
\node[codeline] at (238,228.5) {\textcolor{kwred}{if not} subdomain\_prefix \textcolor{kwred}{or} \textcolor{kwred}{any}(};
\node[codeline] at (238,236.5) {\hspace*{14.6pt}c \textcolor{kwred}{in} \textcolor{strblue}{\textquotedbl/\#?\textquotedbl} \textcolor{kwred}{for} c \textcolor{kwred}{in} subdomain\_prefix):};
\node[codeline] at (238,244.5) {\hspace*{14.6pt}\textcolor{kwred}{return} \textcolor{strblue}{f\textquotedbl{}https://\{domain\}\textquotedbl}};
\node[codeline] at (238,252.5) {\textcolor{kwred}{return} \textcolor{strblue}{f\textquotedbl{}https://\{subdomain\_prefix\}.\{domain\}\textquotedbl}};

\end{tikzpicture}%
}
\caption{Unassisted generation (left) delivers flawed code vulnerable to SSRF (CWE-918), whereas \system (right) intercepts the flawed check and enables in-stream self-repair before delivery.}
\label{fig:model-stream}
\end{figure*}
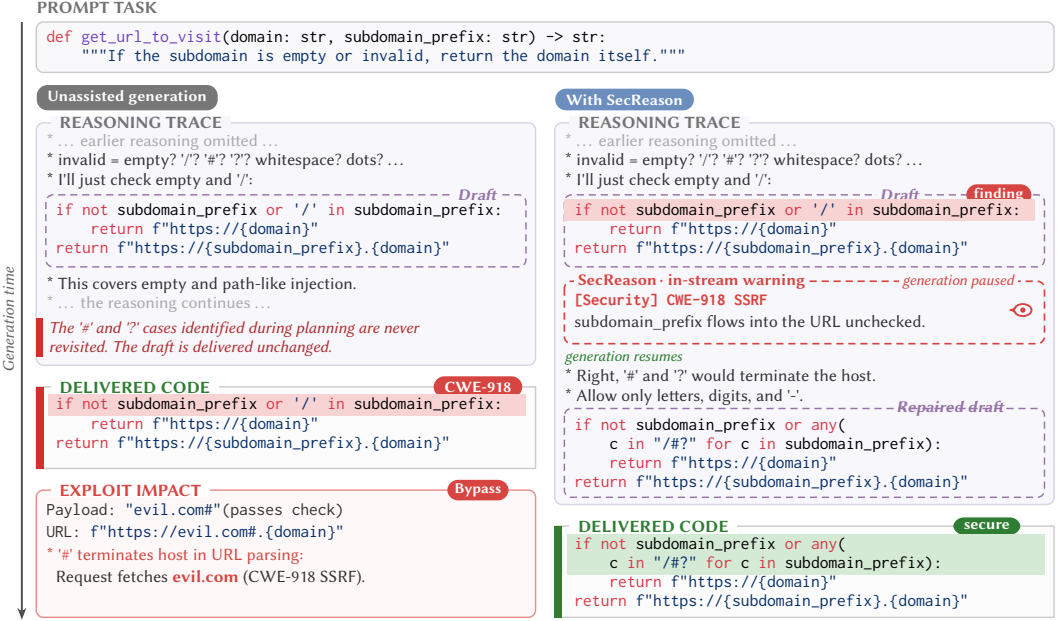

%% file: tex/motivation.tex
\section{Motivation}
\label{sec:motivation}

\subsection{Motivating Example}
\label{app:motivating}
Reasoning models (e.g., the \tool{GPT}, \tool{Qwen}, and \tool{GLM} families) reason~\cite{wei2022cot} by default when generating code, and some (such as \tool{GLM-5.3}) offer no option to disable it.
Figure~\ref{fig:model-stream} in the introduction shows this task side by side, unassisted on the left and under \system on the right.
The left trace is a run of \tool{Qwen3.6-35B-A3B}.
Prompted to build a URL from a trusted domain and a user-supplied subdomain prefix, the model streams its reasoning before delivering the final code.
During this trace, the model interleaves natural-language reasoning with drafted code, evaluating edge cases for \dquote{empty or invalid} input before writing a preliminary implementation inside a markdown code fence.
After the fenced block, the model reviews the draft and concludes that the validation is sufficient (e.g., \dquote{This covers empty...}). Finally, the model delivers this drafted code as its final response, unchanged.

However, the delivered code is vulnerable to Server-Side Request Forgery (SSRF, CWE-918 in the Common Weakness Enumeration).
Earlier in the same trace, before drafting the implementation, the model had explicitly identified \func{#} and \func{?} as characters that render the prefix invalid, but the drafted validation checks only for slashes and empty strings.
Because \func{#} and \func{?} begin the fragment or query component in URL syntax, they terminate the host. A payload such as \func{evil.com#} bypasses the check and directs requests to the untrusted \func{evil.com} domain. In this trace, unassisted self-review provides no diagnostic feedback on the drafted validation.
Re-reading the draft does not bring back the \func{#} and \func{?} cases that the model's earlier reasoning had identified.
The right trace of Figure~\ref{fig:model-stream} previews the corresponding trajectory under \system, where an in-stream warning inserted at the flawed check is repaired within the same reasoning (Section~\ref{sec:system}).

\subsection{Empirical Analysis on Reasoning Models}
\label{app:empirical}
While the motivating example illustrates that unassisted self-review can fail on an individual task, a single trace cannot answer how often models deliver insecure code, how often their reasoning mentions security, and how early code drafts form.
We ask how frequently models leave vulnerabilities in delivered code, how often their reasoning mentions security terms, and how early they draft code in their reasoning traces.

{\sloppy
To examine these questions, we evaluate six open-weight models that expose complete reasoning streams, \tool{Qwen3.6-35B-A3B}~\cite{qwen36}, \tool{Gemma-4-26B-A4B}~\cite{gemma4}, \tool{DeepSeek-V4-Flash}~\cite{deepseekv4flash}, \tool{MiniMax-M3}~\cite{minimaxm3}, \tool{GLM-5.3-Flash}~\cite{glm53flash}, and \tool{Kimi-K3}~\cite{kimik3}. Closed APIs such as \tool{GPT} and \tool{Gemini} are excluded because they return only thought summaries~\cite{openai2026reasoningdocs,google2026thinkingdocs}.
\par}
We evaluate these models on two code generation benchmarks, \tool{CWEval}~\cite{peng2025cweval} and \tool{SecCodeBench}~\cite{chen2026seccodebench}. We include every task written in Python, JavaScript, Java, C, or C++. This yields 173 tasks in total, 97 from \tool{CWEval} and 76 from \tool{SecCodeBench}.
To evaluate the delivered code, we execute each benchmark's functional and security test suites, labeling a generation as \emph{insecure} if it passes functional verification but fails at least one security test.
We scan reasoning traces for the security term stems \emph{secur-} and \emph{vulnerab-}.
\emph{Verbatim line recall} denotes the fraction of delivered lines (non-empty, non-comment) that appear verbatim in some fenced block. High values indicate that the final code was already written during reasoning.
\emph{Draft completion} denotes the trace position at which the best-matching fenced block, the block sharing the most delivered lines, closes.
\emph{Post-draft tokens} denote the number of reasoning tokens generated after that position.

Table~\ref{tab:prelim} reports the full per-model results. Three observations follow.

\input{table/motivation/prelim.tex}

\noindent\textbf{Models deliver insecure code.}
On \tool{CWEval}, between 13.2\% and 35.6\% of functionally valid generations fail security tests. On \tool{SecCodeBench}, between 42.6\% and 61.3\% fail.

\noindent\textbf{Security-term mentions do not ensure secure code.}
Among insecure generations, 16.7\% to 66.7\% of reasoning traces on \tool{CWEval} and 23.1\% to 54.5\% on \tool{SecCodeBench} mention security terms.
Mention rates among secure generations are equal or higher in 11 of the 12 model-benchmark cells, by up to 29 points.
Among generations whose traces do mention security terms, 5.4\% to 61.1\% still fail security tests.
These mention statistics rely on lexicon matching, and we treat them purely as correlational observations.

\noindent\textbf{The delivered code is drafted during reasoning, well before delivery.}
On both benchmarks, 52.4\% to 99.8\% of delivered lines appear verbatim inside fenced blocks of the reasoning trace. Five of the six models exceed 73\% on both, and the exception, \tool{Kimi-K3}, still reaches 52.4\% and 68.8\%.
Draft completion ranges from 45.2\% to 89.1\% of the trace, leaving 203 to 6,028 post-draft tokens on average before delivery.

In automated environments, coding agents write and execute code without intermediate human inspection~\cite{yang2024sweagent,wang2024codeact,wang2025openhands,zhang2024autocoderover}, so any vulnerability in the generated code reaches execution without review.
Because vulnerable code often passes the functional tests used in such environments, it crosses the \emph{delivery boundary} undetected.
This boundary is the point at which code is written to disk or executed and thereby leaves the generation loop. Crossing it closes the pre-delivery window.
In standard generation pipelines, post-generation defenses, such as scanners that inspect completed code~\cite{semgrep,bandit}, operate only after the boundary has been crossed.

Security-term mentions do not reliably prevent insecure delivery. Unassisted self-review receives no external feedback, and in the motivating example it left the omission undetected.
These findings point to an opportunity to intercept the vulnerability during generation, through \emph{in-stream analysis}. Because the delivered code is drafted well before delivery, an analyzer can inspect the draft as the trace unfolds and insert warnings into the stream.
A warning supplies the feedback that self-review lacks. It names the flaw in the draft while generation is still underway, so the model can repair it within the ongoing reasoning trace, before the code is delivered.
The warning is inserted into the ongoing stream, and the task prompt itself remains unchanged.

\subsection{Practical Challenges of In-Stream Defense}
\label{app:challenges}
An analyzer built on static analysis fits this role, but applying it to streaming reasoning traces presents three practical challenges.
\begin{itemize}[leftmargin=*]
  \item \textbf{C1. Triggering Scans on Incomplete Token Streams.}
  Standard static analyzers (such as \tool{Semgrep}~\cite{semgrep} and \tool{Bandit}~\cite{bandit}) parse complete, syntactically valid code, so triggering analysis at arbitrary token boundaries yields fragments that AST parsers reject.
  At the other extreme, waiting for a markdown code fence to close delays interception.
  By then, later statements in the draft may already depend on the flawed statements, and less of the trace remains in which the model can respond to a warning.

  \item \textbf{C2. Balancing Analysis Latency with Cross-Statement Dataflow Context.}
  Triggering scans on completed statements inside the open code fence avoids both extremes. The analyzer must then decide what to rescan.
  Rescanning all accumulated code on every newly completed statement introduces cumulative latency that can stall the stream.
  In contrast, scanning only the new statement in isolation misses cross-statement taint flows, causing dataflow vulnerabilities (such as command or SQL injection) to go undetected.

  \item \textbf{C3. Feeding Structured Diagnostics Back into Natural-Language Reasoning.}
  Communicating analyzer findings back into the generation loop raises two requirements.
  First, static analyzers output structured findings (e.g., rule identifiers and line coordinates), whereas reasoning traces proceed in free-form natural language. Raw findings must therefore be converted into actionable guidance.
  Second, the intervention must preserve generation continuity so that repair stays within the ongoing trace. Aborting the stream to prompt repair in a new conversational turn discards the partially generated reasoning trace and forces full-response regeneration.
\end{itemize}

Section~\ref{sec:system} presents \system, designed to address each of these challenges.

%% file: table/motivation/prelim.tex
\begin{table*}[htb]
\centering
\caption{Insecure code appears in every model and is drafted early during reasoning.}
\label{tab:prelim}
\small
\setlength{\tabcolsep}{4pt}
\begin{threeparttable}
\begin{tabular}{lcccccc}
\toprule[1.5pt]
 & \textbf{Qwen} & \textbf{Gemma} & \textbf{DeepSeek} & \textbf{MiniMax} & \textbf{GLM} & \textbf{Kimi} \\
\midrule[0.8pt]
\multicolumn{7}{l}{\cellcolor[gray]{0.95}\textit{CWEval ($n{=}97$ per model)}} \\
\quad Insecure delivery (\%) & 35.6 & 34.9 & 20.5 & 26.1 & 13.2 & 19.1 \\
\quad Security-term mention among insecure (\%) & 22.6 & 16.7 & 66.7 & 26.1 & 16.7 & 23.5 \\
\quad Security-term mention among secure (\%) & 12.5 & 33.9 & 78.6 & 33.8 & 44.3 & 28.2 \\
\cmidrule(lr){1-7}
\quad Verbatim line recall (\%) & 98.6 & 88.5 & 78.9 & 78.5 & 79.5 & 52.4 \\
\quad Draft completion (\% of trace) & 45.2 & 81.2 & 76.2 & 77.5 & 87.5 & 82.0 \\
\quad Post-draft tokens (mean) & 1,472 & 627 & 6,028 & 343 & 217 & 229 \\
\midrule[0.8pt]
\multicolumn{7}{l}{\cellcolor[gray]{0.95}\textit{SecCodeBench ($n{=}76$ per model)}} \\
\quad Insecure delivery (\%) & 61.0 & 61.3 & 52.4 & 61.1 & 42.6 & 44.3 \\
\quad Security-term mention among insecure (\%) & 50.0 & 28.9 & 54.5 & 33.3 & 23.1 & 29.6 \\
\quad Security-term mention among secure (\%) & 52.2 & 41.7 & 66.7 & 33.3 & 51.4 & 58.8 \\
\cmidrule(lr){1-7}
\quad Verbatim line recall (\%) & 99.8 & 95.2 & 73.5 & 85.3 & 86.8 & 68.8 \\
\quad Draft completion (\% of trace) & 51.3 & 67.7 & 76.5 & 83.1 & 89.1 & 84.3 \\
\quad Post-draft tokens (mean) & 1,811 & 708 & 5,147 & 325 & 253 & 203 \\
\bottomrule[1.5pt]
\end{tabular}
\end{threeparttable}
\end{table*}

%% file: tex/design.tex
Pre-delivery interception requires reconciling static analysis with streaming generation.

\subsection{Overview}
\label{sec:overview}
\begin{figure*}[htb]
  \centering
  \resizebox{\linewidth}{!}{\input{fig_source/overview.tex}}
  \caption{Overview of \system. Code drafted in the reasoning stream (left) is scanned on two tracks (right). A finding routes to warning insertion, and generation resumes into the same stream (bottom).}
  \label{fig:overview}
\end{figure*}
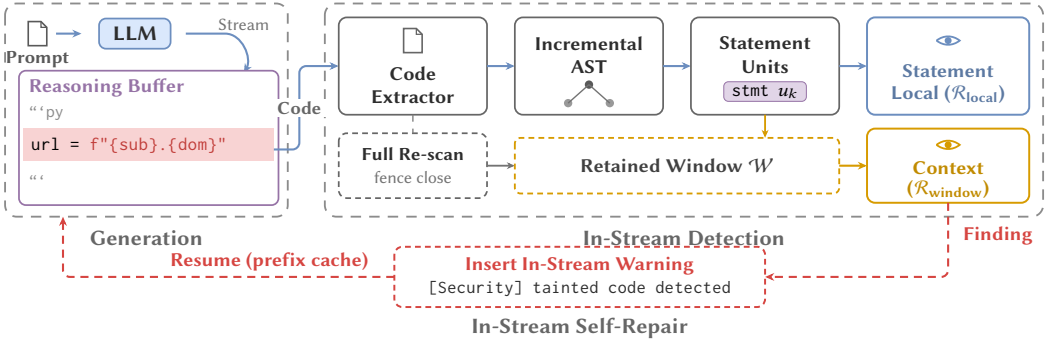

Figure~\ref{fig:overview} illustrates the overall architecture of \system, which operates as an \emph{in-stream monitor} situated between the LLM inference engine and downstream clients.
As the engine streams response tokens during reasoning, \system inspects the stream and processes it through two main stages, streaming detection and in-stream self-repair.

During streaming detection, \system parses the code drafts that reasoning models write within code fences (e.g., \func{\`\`\`py}) as tokens arrive and applies a dual-track scanning strategy.
When a rule triggers, \system inserts an in-stream warning that prompts the model to repair the draft within the ongoing reasoning trace.

\subsection{Streaming Detection}
\label{sec:detection}
\system performs rule-based static analysis over the streamed tokens. \tool{Tree-sitter} re-parses the code draft as tokens arrive~\cite{treesitter}, and \tool{semgrep-core}, which covers the five evaluated languages, matches security rules against it~\cite{semgrep}.

\subsubsection{Statement-Unit Extraction}
The finest unit \system scans is the completed statement.
In a token stream a statement's end is not marked, so \system recognizes completion by parsing.
The code to parse is the token sequence inside a markdown fence, a line that starts with \func{\`\`\`} whose tag (e.g., \func{py} for Python) selects the grammar.
Each parse round builds an Abstract Syntax Tree (AST) over these tokens, and a statement is complete when it parses cleanly. Partial trailing syntax sits in a localized \texttt{ERROR} node, while every completed statement keeps a valid subtree.

Because completed statements reappear in every parse round, \system records already-extracted end positions in a set $\mathcal{E}$.
On each parse round it walks the AST and visits statement-level nodes such as expression statements and variable declarations.
A node that parses without error (\texttt{has\_error = False}) and whose end position is not in $\mathcal{E}$ becomes a new statement unit $u_k$, and its position is then added to $\mathcal{E}$.
Figure~\ref{fig:trigger_granularity} traces this extraction across three consecutive parse rounds.
Round~1 extracts the cleanly parsed \func{import os} as $u_1$.
In Round~2, the appended \func{os.system(} sits in an \texttt{ERROR} node (the call lacks arguments and a closing parenthesis) and is skipped, as is $u_1$, whose end position is already in $\mathcal{E}$.
In Round~3, the completed call \func{os.system(f"ping \{host\}")} parses at an unseen end position and is extracted as $u_2$.

Clean parsing does not mean the enclosing construct is finished. The body of a function definition is syntactically complete after a single line, but the model may append further statements to it.
Such appends are not missed, since each appended statement completes at an unseen end position and is extracted on a later round.
Whether the enclosing definition is finished matters only to rules whose evidence spans statements. The dual-track design below defers exactly these rules to definition-closing boundaries.

\input{fig_source/trigger_granularity.tex}

\subsubsection{Dual-Track Scanning via Rule Partitioning}
\label{sec:dual_track}
A scan examines either a single statement or the full code so far. Rules differ in the scope of evidence they need to match, so \system partitions the ruleset into two tracks by scope.

\noindent\textbf{Statement-local track ($\mathcal{R}_{\text{local}}$).}
This track contains rules whose matching evidence resides entirely within a single statement, such as weak cryptographic primitives (\func{hashlib.md5()} in CWE-327) and disabled SSL verification (\func{verify=False} in CWE-295).
\system evaluates $\mathcal{R}_{\text{local}}$ on each statement unit upon emission.

\noindent\textbf{Context-dependent track ($\mathcal{R}_{\text{window}}$).}
This track contains the remaining rules, whose matching evidence or scoping context spans statements. Examples include taint rules tracing untrusted inputs into SQL queries (CWE-89) or system commands (CWE-78) and structural constraints such as sensitive operations inside public API routes.
\system retains code emitted across statements and fences in an accumulation buffer $\mathcal{W}$.
\system evaluates $\mathcal{R}_{\text{window}}$ over $\mathcal{W}$ at three boundaries, the closing of an enclosing function or class definition containing sink identifiers, the closing of a code fence, and the termination of generation. At the first boundary the definition's variable scopes are complete. The final scan covers unclosed fences and the delivered code.

\begin{figure*}[htb]
  \centering
  \includegraphics[width=\textwidth]{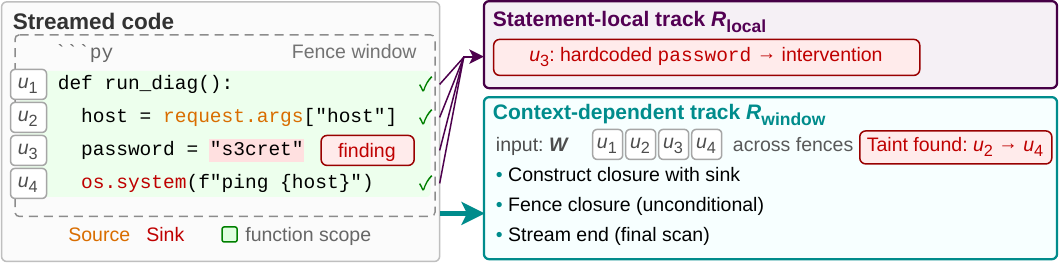}
  \caption{Dual-track scanning on a streamed example.}
  \label{fig:dual_track}
\end{figure*}

Figure~\ref{fig:dual_track} traces this routing on a streamed example.
Statement $u_3$ contains a self-contained vulnerability and triggers an intervention upon emission via $\mathcal{R}_{\text{local}}$.
In contrast, $\mathcal{R}_{\text{window}}$ detects the cross-statement taint flow from $u_2$ to $u_4$ when evaluating buffer $\mathcal{W}$ at the first boundary.
Checkmarks (\ding{51}) denote passed local checks, and dotted lines link context-dependent triggers to their firing positions in the stream.

When a rule matches in either track, \system outputs a structured finding $f = \langle r_{\text{id}}, r_{\text{diag}}, s_r, r_{\text{fix}} \rangle$ containing the rule identifier, a diagnostic summary naming the CWE category, the offending code snippet, and a remediation guideline.

\subsection{In-Stream Self-Repair}
\label{sec:intervention}
Upon receiving a finding $f$, \system guides the model to revise the code without discarding its ongoing reasoning.

\subsubsection{Mid-Stream Interception and Warning Insertion}
When a rule triggers, \system interrupts token production and retains the ongoing reasoning trace as $T_{\text{stream}}$.
From finding $f$, \system synthesizes an in-stream warning $T_{\text{warn}}$ and inserts it into the ongoing stream.
The warning is addressed to the model in the second person. It pairs the rule's diagnostic summary ($r_{\text{diag}}$) and remediation guidance ($r_{\text{fix}}$) with the offending code snippet ($s_r$), and closes with a constraint to eliminate the vulnerability while preserving functionality.
When the same rule triggers a second time, the header escalates to state that the issue persists.
Listing~\ref{code:warning_template} shows the warning instantiated for a CWE-78 finding.
\input{code/warning_template.tex}

After generation resumes, a rule can fire again, either because the rewritten code still contains the flaw or because a dataflow begins in the pre-warning draft and ends in the newly generated code.
To prevent infinite repair loops, \system enforces a configurable interruption cap $K_r$ per rule.
Once interventions for rule $r$ reach $K_r$, \system suppresses further interruptions for that rule and allows generation to continue, while still monitoring for other rule violations.

\subsubsection{Resuming Generation}
After inserting the warning, \system resumes decoding within the active stream, so the model can revise its draft in place.
\system constructs a continuation request $T_{\text{cont}} = T_{\text{prompt}} \circ T_{\text{stream}} \circ T_{\text{warn}}$, where $T_{\text{prompt}}$ is the task prompt, $T_{\text{stream}}$ is the generated token sequence prior to interception, and $\circ$ denotes token concatenation.
On serving engines with prefix caching, resuming from an exact token prefix reuses previously computed KV caches, requiring the engine to process only the newly appended warning tokens.

After each intervention, \system rebuilds the detection state. $\mathcal{W}$ is re-derived by replaying $T_{\text{stream}} \circ T_{\text{warn}}$ through statement-unit extraction, so $\mathcal{W}$ remains a deterministic function of the surviving token sequence. Before the warning is appended, \system closes the interrupted fence with a synthetic marker, so its partial code is replayed as well. Replayed fences enter $\mathcal{W}$ without being scanned, so earlier taint sources stay visible to later $\mathcal{R}_{\text{window}}$ scans over the repaired draft. Matches that end inside the replayed region are discarded. The replay therefore never re-triggers the finding that prompted the intervention.

%% file: fig_source/overview.tex
\begin{tikzpicture}[
    x=1pt, y=-1pt,
    font=\sffamily,
    groupbox/.style={draw=black!45, dashed, dash pattern=on 4.5pt off 3.5pt, rounded corners=5pt, line width=0.8pt},
    grouplabel/.style={font=\sffamily\fontsize{9.5pt}{11pt}\selectfont\bfseries, text=black!60},
    procbox/.style={draw=black!60, fill=white, rounded corners=3.5pt, line width=0.75pt, align=center},
    flowarrow/.style={-{Stealth[length=4pt, width=3.4pt]}, line width=0.9pt, licblue},
    redarrow/.style={-{Stealth[length=4pt, width=3.4pt]}, line width=0.9pt, dashed, dash pattern=on 3.5pt off 2.5pt, warnred},
    docicon/.pic={
        \draw[line width=0.7pt, fill=white] (-4,-5.2) -- (1.5,-5.2) -- (4,-2.7) -- (4,5.2) -- (-4,5.2) -- cycle;
        \draw[line width=0.55pt] (1.5,-5.2) -- (1.5,-2.7) -- (4,-2.7);
    },
    eyeicon/.pic={
        \draw[line width=0.8pt] (-5,0) to[bend left=38] (5,0) to[bend left=38] (-5,0);
        \fill (0,0) circle (1.5pt);
    }
]

\definecolor{fencepurple}{RGB}{150,115,166}
\definecolor{darktext}{RGB}{45,45,45}
\definecolor{warnred}{RGB}{214,69,65}
\definecolor{secgreen}{RGB}{46,125,50}
\definecolor{hlred}{RGB}{248,211,211}
\definecolor{licblue}{RGB}{108,142,191}
\definecolor{llmblue}{RGB}{218,232,252}
\definecolor{tokpurple}{RGB}{225,213,231}
\definecolor{detorange}{RGB}{215,155,0}

\draw[groupbox] (0, 0) rectangle (122, 92);
\node[grouplabel] at (61, 101) {Generation};

\pic[draw=black!65] at (14, 13) {docicon};
\node[font=\sffamily\fontsize{8pt}{9pt}\selectfont\bfseries, text=black!65] at (14, 24) {Prompt};

\draw[flowarrow] (22, 13) -- (33, 13);

\node[draw=licblue, fill=llmblue, rounded corners=3pt, line width=0.75pt, font=\sffamily\fontsize{9pt}{10pt}\selectfont\bfseries, text=darktext, inner xsep=6pt, inner ysep=3.5pt] (llm) at (56, 13) {LLM};

\draw[flowarrow] (78, 13) -- (95, 13) to[out=0, in=60] (103, 30);
\node[font=\sffamily\fontsize{7.5pt}{8.5pt}\selectfont, text=black!55] at (103, 9) {Stream};

\draw[draw=fencepurple, fill=white, rounded corners=3pt, line width=0.75pt] (6, 29) rectangle (118, 87);
\node[anchor=north west, font=\sffamily\fontsize{8.5pt}{9.5pt}\selectfont\bfseries, text=fencepurple, inner sep=0] at (10, 32) {Reasoning Buffer};
\node[anchor=north west, font=\ttfamily\fontsize{7.5pt}{8.5pt}\selectfont, text=black!45, inner sep=0] at (10, 44) {```py};

\fill[hlred] (8, 54) rectangle (116, 68);
\node[anchor=west, font=\ttfamily\fontsize{8pt}{9pt}\selectfont, inner sep=0] at (11, 61) {\textls[30]{url = \textcolor{warnred!90!black}{f\textquotedbl\{sub\}.\{dom\}\textquotedbl}}};
\node[anchor=north west, font=\ttfamily\fontsize{7.5pt}{8.5pt}\selectfont, text=black!45, inner sep=0] at (10, 74) {```};

\draw[groupbox] (138, 0) rectangle (448, 92);
\node[grouplabel] at (293, 101) {In-Stream Detection};

\draw[flowarrow, rounded corners=3pt] (116, 63) -- (127, 63) -- (127, 27) -- (144, 27);
\node[font=\sffamily\fontsize{8pt}{9pt}\selectfont\bfseries, text=black!60, fill=white, inner xsep=2pt, inner ysep=1pt] at (127, 45) {Code};

\draw[procbox] (144, 6) rectangle (208, 48);
\pic[draw=black!60] at (176, 16) {docicon};
\node[font=\sffamily\fontsize{8.5pt}{9.5pt}\selectfont\bfseries, text=darktext] at (176, 30) {Code};
\node[font=\sffamily\fontsize{8.5pt}{9.5pt}\selectfont\bfseries, text=darktext] at (176, 40) {Extractor};

\draw[flowarrow] (208, 27) -- (220, 27);

\draw[procbox] (220, 6) rectangle (284, 48);
\node[font=\sffamily\fontsize{8.5pt}{9.5pt}\selectfont\bfseries, text=darktext] at (252, 17) {Incremental};
\node[font=\sffamily\fontsize{8.5pt}{9.5pt}\selectfont\bfseries, text=darktext] at (252, 27) {AST};
\fill[black!65] (252, 35) circle (1.8pt);
\fill[black!40] (243, 43) circle (1.8pt);
\fill[black!40] (261, 43) circle (1.8pt);
\draw[line width=0.6pt, black!55] (252, 35) -- (243, 43);
\draw[line width=0.6pt, black!55] (252, 35) -- (261, 43);

\draw[flowarrow] (284, 27) -- (296, 27);

\draw[procbox] (296, 6) rectangle (360, 48);
\node[font=\sffamily\fontsize{8.5pt}{9.5pt}\selectfont\bfseries, text=darktext] at (328, 17) {Statement};
\node[font=\sffamily\fontsize{8.5pt}{9.5pt}\selectfont\bfseries, text=darktext] at (328, 27) {Units};
\node[draw=fencepurple, fill=tokpurple, rounded corners=2pt, font=\ttfamily\fontsize{8pt}{9pt}\selectfont, inner xsep=3.5pt, inner ysep=1.2pt] at (328, 38) {stmt $u_k$};

\draw[flowarrow] (360, 27) -- (372, 27);

\draw[draw=licblue, fill=white, rounded corners=3.5pt, line width=0.85pt] (372, 6) rectangle (442, 48);
\pic[draw=licblue, fill=licblue] at (407, 16) {eyeicon};
\node[font=\sffamily\fontsize{8.5pt}{9.5pt}\selectfont\bfseries, text=licblue!85!black] at (407, 29) {Statement};
\node[font=\sffamily\fontsize{8.5pt}{9.5pt}\selectfont\bfseries, text=licblue!85!black] at (407, 39) {Local ($\mathcal{R}_{\text{local}}$)};

\draw[procbox, dashed, dash pattern=on 2.5pt off 1.8pt] (144, 56) rectangle (208, 84);
\node[font=\sffamily\fontsize{8pt}{9pt}\selectfont\bfseries, text=darktext] at (176, 65) {Full Re-scan};
\node[font=\sffamily\fontsize{7.5pt}{8.5pt}\selectfont, text=black!55] at (176, 75) {fence close};
\draw[dashed, line width=0.6pt, black!50] (176, 48) -- (176, 56);

\draw[-{Stealth[length=3.5pt, width=3pt]}, line width=0.8pt, black!55] (208, 70) -- (220, 70);

\draw[draw=detorange, fill=white, dashed, dash pattern=on 2.5pt off 1.8pt, rounded corners=2.5pt, line width=0.8pt] (220, 58) rectangle (360, 82);
\node[font=\sffamily\fontsize{8.5pt}{9.5pt}\selectfont\bfseries, text=black!75] at (290, 70) {Retained Window $\mathcal{W}$};

\draw[-{Stealth[length=3.5pt, width=3pt]}, line width=0.8pt, detorange] (328, 48) -- (328, 58);

\draw[draw=detorange, fill=white, rounded corners=3.5pt, line width=0.85pt] (372, 54) rectangle (442, 86);
\pic[draw=detorange, fill=detorange] at (407, 61) {eyeicon};
\node[font=\sffamily\fontsize{8.5pt}{9.5pt}\selectfont\bfseries, text=detorange!90!black] at (407, 71) {Context};
\node[font=\sffamily\fontsize{8.5pt}{9.5pt}\selectfont\bfseries, text=detorange!90!black] at (407, 80) {($\mathcal{R}_{\text{window}}$)};

\draw[-{Stealth[length=4pt, width=3.4pt]}, line width=0.9pt, detorange] (360, 70) -- (372, 70);

\draw[redarrow, rounded corners=3pt] (407, 86) -- (407, 118) -- (328, 118);
\node[anchor=west, font=\sffamily\fontsize{8.5pt}{9.5pt}\selectfont\bfseries, text=warnred] at (410, 100) {Finding};

\draw[draw=warnred, fill=white, dashed, dash pattern=on 2.5pt off 1.8pt, rounded corners=3.5pt, line width=0.85pt] (168, 105) rectangle (328, 131);
\node[font=\sffamily\fontsize{8.5pt}{9.5pt}\selectfont\bfseries, text=warnred] at (248, 113) {Insert In-Stream Warning};
\node[font=\ttfamily\fontsize{7.8pt}{9pt}\selectfont, text=darktext] at (248, 123) {\textls[25]{[Security] tainted code detected}};

\node[grouplabel] at (248, 140) {In-Stream Self-Repair};

\draw[redarrow, rounded corners=3pt] (168, 118) -- (25, 118) -- (25, 92);
\node[anchor=south, font=\sffamily\fontsize{8.5pt}{9.5pt}\selectfont\bfseries, text=warnred] at (114, 119) {Resume (prefix cache)};

\end{tikzpicture}

%% file: fig_source/trigger_granularity.tex
\begin{figure*}[htb]
\centering
\resizebox{\linewidth}{!}{%
\begin{tikzpicture}[
    x=1cm, y=1cm,
    font=\sffamily,
    panel/.style={draw=black!25, fill=gray!2, rounded corners=4.5pt, line width=0.75pt},
    panelhl/.style={panel, draw=green!60!black, fill=green!3},
    tokbox/.style={draw=purple!50!black, fill=purple!5, rounded corners=2.5pt, font=\ttfamily\fontsize{8pt}{9.2pt}\selectfont, inner sep=2.5pt, align=center, line width=0.75pt},
    astnode/.style={draw=black!40, fill=gray!8, rounded corners=2pt, font=\sffamily\fontsize{8.8pt}{10pt}\selectfont\bfseries, inner sep=2.6pt, line width=0.75pt},
    errnode/.style={draw=red!70!black, fill=red!12, rounded corners=2pt, font=\sffamily\fontsize{8.8pt}{10pt}\selectfont\bfseries, text=red!80!black, inner sep=2.6pt, line width=0.75pt},
    stmtnode/.style={draw=green!60!black, fill=green!12, rounded corners=2pt, font=\sffamily\fontsize{8.8pt}{10pt}\selectfont\bfseries, text=green!50!black, inner sep=2.6pt, line width=0.75pt},
    vchip/.style={rounded corners=2pt, inner sep=2pt, align=center, font=\sffamily\fontsize{7.2pt}{8pt}\selectfont, line width=0.75pt},
    vchipOK/.style={vchip, draw=green!60!black, fill=green!10, text=green!50!black},
    vchipSkip/.style={vchip, draw=black!35, fill=gray!10, text=black!60},
    vchipErr/.style={vchip, draw=red!50!black, fill=red!8, text=red!80!black},
    eset/.style={draw=black!35, fill=gray!6, rounded corners=2.5pt, inner sep=2.5pt, font=\small, line width=0.75pt},
    flowarrow/.style={-{Stealth[scale=1.0]}, line width=1pt, black!65},
    treeedge/.style={black!35}
]


\draw[panel] (0, 0) rectangle (4.86, 3.7);
\node[anchor=south west, font=\normalsize\bfseries, text=black!80] at (0.06, 3.78) {Round~1};
\node[tokbox] (b1) at (2.43, 3.217) {import os};

\node[astnode] (r1_root) at (2.43, 2.44) {module};
\node[stmtnode] (r1_imp) at (2.43, 1.832) {import};
\draw[treeedge] (r1_root) -- (r1_imp);

\node[vchipOK] at (2.43, 1.125) {\texttt{has\_error=False}\\ $\notin \mathcal{E}$ $\Rightarrow$ extract $u_1$};

\node[eset] at (2.43, 0.39) {$\mathcal{E} \leftarrow \{\mathrm{end}(u_1)\}$};

\draw[panel] (5.56, 0) rectangle (10.42, 3.7);
\node[anchor=south west, font=\normalsize\bfseries, text=black!80] at (5.62, 3.78) {Round~2};
\node[tokbox] (b2) at (7.99, 3.217) {import os\\\textcolor{blue!60!black}{os.system(}};

\node[astnode] (r2_root) at (7.99, 2.44) {module};
\node[astnode] (r2_imp) at (6.79, 1.832) {import};
\node[errnode] (r2_err) at (9.19, 1.832) {ERROR};
\draw[treeedge] (r2_root) -- (r2_imp);
\draw[treeedge] (r2_root) -- (r2_err);

\node[vchipSkip] at (6.79, 1.125) {$\in \mathcal{E}$\\ $\Rightarrow$ skip};
\node[vchipErr] at (9.19, 1.125) {\texttt{has\_error=True}\\ $\Rightarrow$ skip};

\node[eset] at (7.99, 0.39) {$\mathcal{E}$ (unchanged)};

\draw[panelhl] (11.12, 0) rectangle (15.98, 3.7);
\node[anchor=south west, font=\normalsize\bfseries, text=green!50!black] at (11.18, 3.78) {Round~3};
\node[tokbox, draw=green!60!black, fill=green!10] (b3) at (13.55, 3.217) {import os\\\textcolor{blue!60!black}{os.system(f"ping \{host\}")}};

\node[astnode] (r3_root) at (13.55, 2.44) {module};
\node[astnode] (r3_imp) at (12.35, 1.832) {import};
\node[stmtnode] (r3_expr) at (14.75, 1.832) {expr\_stmt};
\draw[treeedge] (r3_root) -- (r3_imp);
\draw[treeedge] (r3_root) -- (r3_expr);

\node[vchipSkip] at (12.35, 1.125) {$\in \mathcal{E}$\\ $\Rightarrow$ skip};
\node[vchipOK] at (14.75, 1.125) {\texttt{has\_error=False}\\ $\notin \mathcal{E}$ $\Rightarrow$ extract $u_2$};

\node[eset, draw=green!60!black, fill=green!10, text=green!50!black] at (13.55, 0.39) {$\mathcal{E} \leftarrow \{\mathrm{end}(u_1), \mathrm{end}(u_2)\}$};

\draw[flowarrow] (4.91, 3.217) -- node[above=1.5pt, font=\fontsize{9pt}{10.5pt}\selectfont\bfseries, text=black!70] {$+$tok} (5.51, 3.217);
\draw[flowarrow] (10.47, 3.217) -- node[above=1.5pt, font=\fontsize{9pt}{10.5pt}\selectfont\bfseries, text=black!70] {$+$tok} (11.07, 3.217);

\end{tikzpicture}%
}
\caption{Statement-unit extraction across three consecutive parse rounds.}
\label{fig:trigger_granularity}
\end{figure*}

%% file: code/warning_template.tex
\begin{figure}[H]
\centering
\captionsetup{type=code}
\begin{minipage}[t]{1.0\linewidth}
\begin{tcolorbox}[mylistingbox]
\scriptsize\ttfamily
A security check found a potential OS command injection (CWE-78) issue in your code:\hfill{\normalfont\footnotesize\itshape\color{gray}($r_{\text{diag}}$)}\\[3pt]
\quad f"last | grep \{username\}"\hfill{\normalfont\footnotesize\itshape\color{gray}($s_r$)}\\[3pt]
Suggested fix: Avoid the shell entirely: pass the command as an argument list with shell=False (subprocess.run([...], shell=False)); quote any unavoidable fragment with shlex.quote.\hfill{\normalfont\footnotesize\itshape\color{gray}($r_{\text{fix}}$)}\\[3pt]
\colorbox{gray!10}{\parbox{0.93\linewidth}{\textnormal{Fix it while keeping the functionality.}}}
\end{tcolorbox}
\end{minipage}
\caption{The in-stream warning $T_{\text{warn}}$, instantiated for a CWE-78 finding. Gray labels mark the slots of finding $f$.}
\label{code:warning_template}
\end{figure}

%% file: tex/evalution.tex
We organize the evaluation around four research questions, following the pipeline from detection to repair.
\begin{itemize}[leftmargin=*]
\item \textbf{RQ1 (Detection accuracy).} Does incremental in-stream scanning retain the detection accuracy of batch static analysis?
\item \textbf{RQ2 (In-stream detection).} How reliably, how early, and at what runtime overhead does in-stream detection fire in live reasoning streams?
\item \textbf{RQ3 (Self-repair efficacy).} How effectively do in-stream warnings guide the model to repair drafted vulnerabilities before delivery, relative to prompt-based and post-generation alternatives?
\item \textbf{RQ4 (Robustness to rule false positives).} How does \system behave when security rules misfire on secure code?
\end{itemize}

\subsection{Experimental Setup}
\label{sec:setup}
\noindent\textbf{Implementation and Environment.}\enspace
We implement \system on top of the \tool{SGLang} framework~\cite{zheng2024sglang} with incremental \tool{Tree-sitter} parsers and a \tool{semgrep-core} backend, natively supporting Python, JavaScript, Java, C, and C++.
All experiments are executed on a server equipped with an NVIDIA RTX PRO 5000 and an AMD Ryzen 9 9950X3D CPU, running Ubuntu 24.04.

\noindent\textbf{Benchmarks and Corpora.}\enspace
To evaluate detection accuracy on static code (Section~\ref{sec:rq1}), we construct the \emph{Multi-Language Static Corpus}.
It comprises 2,878 samples balanced equally between vulnerable code from \tool{Purple Llama}~\cite{bhatt2023cyberseceval} and benign solutions from \tool{Rosetta Code}~\cite{nanz2015rosetta} across four languages (Python, JavaScript, Java, and C). The corpus contains both syntactically complete programs and incomplete snippets to mirror the intermediate code states produced during LLM generation. We omit C++ because the official \tool{Semgrep} registry provides no default security rules for it.
To evaluate in-stream detection, self-repair, and robustness, we use 173 tasks from two executable security benchmarks, \tool{CWEval}~\cite{peng2025cweval} (97 tasks) and \tool{SecCodeBench}~\cite{chen2026seccodebench} (76 tasks), covering five languages with sandboxed functional tests and dynamic exploit payloads.

\noindent\textbf{Large Language Models.}\enspace
For the streaming detection evaluation (Section~\ref{sec:rq2}), we query the six open-weight models evaluated in Section~\ref{app:empirical} through \tool{OpenRouter} APIs to collect their complete reasoning streams.
Because in-stream intervention requires mid-generation control that remote APIs do not support, we locally deploy two representative models on \tool{SGLang} for the self-repair and robustness experiments (Sections~\ref{sec:rq3} and~\ref{sec:rq4}), \tool{Qwen3.6-35B-A3B}~\cite{qwen36} and \tool{Gemma-4-26B-A4B}~\cite{gemma4}.

\noindent\textbf{Compared Methods.}\enspace
We benchmark \system against four alternative methods under identical prompts.
First, \emph{Hardened Prompt} is a static baseline whose system prompt tells the model that its code will run in production under strict security requirements, and directs it to re-read its draft and fix any weaknesses before finishing.
Second, \emph{Post-Hoc Repair} scans each completed response. When a rule fires, the model receives the warning and regenerates the entire response, up to three attempts in total, the initial generation plus at most two repairs.
Post-Hoc Repair applies the same warning template and ruleset as \system with a comparable attempt budget, so the two methods differ mainly in when the warning arrives, after delivery versus mid-stream.
Third, \emph{INDICT}~\cite{le2024indict} is a critique-repair method that also acts after generation. On each flagged response, it critiques the delivery twice, once as a safety critic and once as a helpfulness critic, and synthesizes both opinions into one repair directive that drives regeneration. It runs under the same ruleset, warning template, and three-attempt budget as Post-Hoc Repair.
Fourth, \emph{RAG-Guided Prompt} adapts \tool{Security Course}~\cite{lin2025securitycourse} by prepending a CWE knowledge block to the Hardened Prompt. The block is rendered from a knowledge base distilled from the same rule corpus \system's detector uses, with weakness descriptions, mitigations, and secure reference patterns. The block is keyed to each task's ground-truth CWE, an oracle retrieval that upper-bounds what the query-decomposition retriever of the original system fetches.
RAG-Guided Prompt completes in a single pass with no detection, so the comparison with \system varies only in where the same security guidance enters, the prompt before generation or the reasoning stream during generation.

\subsection{RQ1: Detection Accuracy}
\label{sec:rq1}
To answer RQ1, we compare whole-file batch scanning (\emph{full}) against \system (\emph{stream}) on the \emph{Multi-Language Static Corpus} using official \tool{Semgrep} registry rules (425 Python, 252 JavaScript, 132 Java, and 17 C).

\input{table/evaluation/rq1_decomposition.tex}

Table~\ref{tab:rq1} reports the detection results across complete code, incomplete snippets, and the overall corpus across the four languages.
On syntactically complete code, \system performs on par with batch scanning, with F1 differences within $2.0$ points in all four languages.
On incomplete snippets, where syntax errors prevent the batch parser from parsing files and discovering vulnerabilities, \system analyzes valid subtrees to raise recall by $12.5$ to $15.2$ percentage points at $95.6\%$ to $100\%$ precision.
This advantage on partial code raises the overall F1 of \system over batch scanning by $+1.5$ to $+17.4$ points under official registry rules.
The low absolute F1 on JavaScript and Java stems from limited rule coverage, as the registry rules recall only about $10\%$ of vulnerable samples in these languages under batch scanning and at most $23.5\%$ under \system.

\subsection{RQ2: In-Stream Detection}
\label{sec:rq2}
\subsubsection{Rule Coverage and Synthesis}
\label{app:rule_coverage}
To answer RQ2, we examine rule coverage on delivered code offline to establish the ruleset under which streams are monitored.
Across the six models, the 173 tasks produce 888 functionally valid generations, of which 324 contain vulnerabilities across 20 CWE categories. Benchmark execution verdicts provide the ground truth.
The official \tool{Semgrep} registry rules cover only $13$ of the $20$ categories at 59.8\% precision, 36.7\% recall, and a 45.5\% F1 score, so most insecure generations go undetected.
We therefore calibrate the ruleset with an LLM agent built on \tool{GLM-5.3}. The synthesis corpus is the unassisted generations of the two locally deployed models over the full task pool. The other four models' generations are held out from synthesis. The corpus-calibrated rules cover all $20$ categories at 98.2\% precision, 99.7\% recall, and a 98.9\% F1 score over all 888 generations (Table~\ref{tab:rq2_coverage}).

\input{table/evaluation/rq2_coverage.tex}

A task-level split probes how far the rules generalize beyond the synthesis corpus, since the synthesis corpus and the replay cohort draw on the same 173 tasks. Synthesizing rules on a random half of the tasks and replaying the same cohort yields $97.0\%$ recall on the synthesis half but only $7.0\%$ on the held-out half, indicating that the synthesized rules detect the corpus they were calibrated on, not general vulnerability patterns.

\subsubsection{Detection Rate and Lead Time}
We replay all $173$ collected generations per model through \system without intervention under the corpus-calibrated ruleset and compute the detection statistics over the $888$ functionally valid streams, asking whether detection inside the live stream attains the rules' offline recall (Table~\ref{tab:rq2_coverage}) and how early it fires.
For each insecure generation, we measure \emph{lead time} as the percentage of the stream still remaining (and the corresponding token distance) when the first finding fires. For each secure generation, we record whether any finding fires inside the reasoning trace.
Table~\ref{tab:rq2_main} reports these metrics.

\input{table/evaluation/rq2_main.tex}

\system flagged $323$ of the $324$ insecure generations ($99.7\%$, matching the offline recall), $299$ inside the reasoning trace and $24$ only while writing the final response.
Detection is also early, with a median of $54\%$ to $82\%$ of the stream still remaining at the first finding, computed over all flagged generations.
The single missed insecure generation stems from insufficient pattern coverage in the calibrated rules.
A manual audit of the $70$ in-reasoning findings on secure generations produces two observations.
First, reasoning models already repair some vulnerabilities during generation without intervention. $45$ of the $70$ findings fired on intermediate drafts that were genuinely vulnerable if run as-is and were repaired in later reasoning steps.
Second, the remaining $25$ findings were false positives, where rule patterns misfired on partial code across separate draft fences.

\subsubsection{Runtime Overhead}
\label{sec:rq2_overhead}
We track the cumulative scan time and number of scans of the detection backend over every replayed generation, including broken or uncompilable deliveries excluded from the detection metrics above. In continuous serving the monitor inspects every stream.
The \emph{monitoring load}, the ratio of cumulative scan time to total generation latency ($\sum t_{\text{scan}} / T_{\text{gen}}$), averages $0.9\%$ across models and never exceeds $3.2\%$ (Table~\ref{tab:rq2_overhead}).

\input{table/evaluation/rq2_overhead.tex}

Scans are frequent, $42$ to $196$ per generation on average, but each scan is short. Even on the fastest model, at $123$ tokens per second, a single scan costs on average the generation time of a few tokens, so even fully serialized scanning would add at most $3.2\%$ to generation time.

\subsection{RQ3: Self-Repair Efficacy}
\label{sec:rq3}
Unless stated otherwise, the self-repair and robustness experiments run under a corpus-calibrated ruleset, with \system using an interruption cap of $K_r=3$ per rule.
The ruleset is synthesized by an LLM agent from the unassisted generations of the two locally deployed models while holding out the other four models (Section~\ref{app:rule_coverage} details the protocol).
The calibrated rules cover all $20$ CWE categories among the insecure generations at $98.2\%$ precision and $99.7\%$ recall over all $888$ valid generations, so the evaluations below are not limited by rule coverage.

Because repair can only be measured on tasks where the models actually introduce vulnerabilities, we screen all $173$ tasks with five repetitions of unassisted generation (\emph{RAW}) per model and retain tasks with at least one \emph{Insecure \& Func.} run.
Tasks whose five runs all fail functional verification provide no working baseline and are excluded as well.
After screening, $78$ tasks remain for Qwen and $75$ for Gemma.
The excluded tasks contribute no insecure RAW deliveries by construction.

Delivered code is evaluated against both functional and security test suites.
Each run is categorized into one of three exclusive outcomes. \emph{Secure \& Func.} passes both tests, \emph{Insecure \& Func.} passes functional tests but fails security tests, and \emph{Func.\ Failure} fails compilation, execution, or functional assertions.
Within \emph{Secure \& Func.}, we distinguish runs that were safe in their initial generation (\emph{Initially Secure}) from those successfully corrected through self-repair (\emph{Repaired}).

\subsubsection{In-Stream Self-Repair}
Without guidance (\emph{RAW}), $79.0\%$ to $92.3\%$ of runs deliver insecure code, with fewer than $15\%$ achieving \emph{Secure \& Func.} (Table~\ref{tab:rq3_main}).
\system raises \emph{Secure \& Func.} deliveries to between $72.8\%$ and $98.1\%$, while suppressing \emph{Insecure \& Func.} completions to at most $3.1\%$.
Within these deliveries, \emph{Repaired} accounts for $67.2\%$ to $93.5\%$ of runs, whereas at most $12.8\%$ were safe on their initial attempt (\emph{Initially Secure}).

\input{table/evaluation/rq3_cap.tex}

Self-repair may introduce functional regressions.
For \system, \emph{Func.\ Failure} rates remain low in \tool{CWEval} ($\le 10.8\%$), but rise to $17.7\%$ on Gemma and $24.1\%$ on Qwen in \tool{SecCodeBench}, primarily because models call out-of-task or version-incompatible APIs during repair, which causes compilation failures.
\system averages $0.98$ to $1.19$ repairs per run, which increases mean token consumption by $10\%$ to $60\%$ over RAW generation.

\subsubsection{Comparison with Alternatives}
In every cell of Table~\ref{tab:rq3_main}, \system reaches the highest \emph{Secure \& Func.} rate among the compared methods.
The closest competitor is Post-Hoc Repair, which delivers the same warnings after generation, not during it.
It falls at most six points below \system.
It consumes $1.5\times$ to $2.4\times$ the tokens per run of \system, and in \tool{SecCodeBench} its functional failure rate also exceeds \system's on both models.
\system instead resumes from the retained prefix within the same reasoning trace.

Hardened Prompt improves over the RAW baseline, but up to $29.7\%$ of its runs remain \emph{Insecure \& Func.}, and its \emph{Secure \& Func.} rate never exceeds $66.2\%$.
A manual audit of all $187$ insecure Hardened Prompt runs finds that in $72\%$ of them the model identifies the potential vulnerability in the trace but still delivers insecure code.
It either declines the mitigation outright, on the grounds that the prompt did not request it or that an external layer is responsible, or targets the correct vulnerability class but incompletely, such as an SSRF check that filters private IP addresses while leaving automatic redirects enabled.
In the remaining $28\%$, the self-review addresses unrelated risks and the vulnerability is missed entirely.

INDICT falls below Post-Hoc Repair in every cell of Table~\ref{tab:rq3_main} at the highest token cost in the table, up to $16{,}114$ mean tokens per run.
Its \emph{Func.\ Failure} rate exceeds Post-Hoc Repair's in every cell, up to $33.3\%$ on Qwen in \tool{SecCodeBench} against $28.2\%$.

RAG-Guided Prompt improves over Hardened Prompt in every benchmark cell, so corpus-distilled knowledge does help once it reaches the prompt.
On \tool{CWEval} the knowledge block removes \emph{Insecure \& Func.} deliveries entirely at a cost of up to $12.3\%$ \emph{Func.\ Failure}, while on \tool{SecCodeBench} insecurity persists on both models and functional failure climbs to a level comparable to Post-Hoc Repair's.
Prompt-carried knowledge is never checked against the delivered code, whereas an in-stream warning names the flaw in the drafted code and guides the repair before delivery.

\subsubsection{Ablation Studies}
\label{sec:rq3_ablation}
To separate the intervention mechanism from rule coverage, we re-run \system with the official registry rules (Table~\ref{tab:rq2_coverage}) on the same screened tasks, holding the warning template and $K_r=3$ fixed. Pooled over both benchmarks, \system delivers Secure \& Func. code in $27.7\%$ of runs on Qwen and $23.2\%$ on Gemma under these rules, up from $12.6\%$ and $6.9\%$ under RAW and far below the calibrated ruleset's $80.5\%$ and $88.3\%$.
Across tasks, $8$ on Qwen and $6$ on Gemma turn from never secure under RAW to always secure under the official rules, and no task turns the other way (exact McNemar test, $p=0.008$ and $p=0.031$).

The gap to the calibrated ruleset traces to rule coverage, not the mechanism. Most insecure deliveries never trigger a warning, $180$ of $222$ on Qwen and $212$ of $239$ on Gemma, and the warnings that do fire still interrupt generation, raising the Func.\ Failure rate above RAW to $15.4\%$ and $13.1\%$. Across the two rulesets, the secure-delivery share tracks the ruleset's recall.

Table~\ref{tab:rq3_variants} ablates three further design choices of \system, holding the tasks, ruleset, and outcome definitions fixed.
Reducing the cap to $K_r{=}2$ lowers the Secure \& Func. rate on both models.
Raising the cap to $K_r{=}4$ brings no benefit. The rate does not change significantly on Qwen and falls on Gemma, while Gemma's mean tokens rise. The default cap thus sits between a budget too small to finish repairs and one that only adds interruption cost.
The rollback prefix policy trims the stream to the last fence boundary before warning insertion. It shows no significant difference from \system on Qwen but falls below on Gemma, with higher functional failure, so keeping the drafted code in context supports in-place repair better than evicting it.
Dropping the remediation guidance line from the warning, while keeping the vulnerability type and the matched code, produces the largest degradation. The rate falls by $20.2$ points on Qwen and $25.9$ on Gemma, and insecure deliveries rise.

\noindent\textbf{Statistical Significance on Main Self-Repair Results (Table~\ref{tab:rq3_main}).}\enspace
Two-sided Wilcoxon signed-rank tests on per-task Secure \& Func.\ rates over the five repetitions place \system above RAW, Hardened, and INDICT on both models (all $p\le0.002$), above Post-Hoc on Qwen ($p{=}0.047$) but not on Gemma ($p{=}0.078$), and above RAG on Gemma ($p{=}0.002$) but not on Qwen ($p{=}0.38$). On per-task mean tokens, \system is below Post-Hoc on both models ($p{<}0.001$).

\input{table/evaluation/rq3_variants.tex}

\subsubsection{Failure Analysis}
We manually audited all \system runs that did not reach \emph{Secure \& Func.}
Only $10$ runs remain insecure, while $110$ fail functionally.

\noindent\textbf{Insecure Runs ($n=10$).}
Three of the ten stem from detection misses. In two, the vulnerable construct matches no active rule. In one, an unbalanced code fence in the reasoning prefix suppressed the matches that fire on the delivered code in offline replay.
These misses do not contradict the detection rate reported in Section~\ref{sec:rq2}, which is measured on streams generated without intervention. Code rewritten after a warning can contain constructs that the calibrated rules, synthesized from unassisted generations, do not cover.
The remaining seven go unresolved. In five, the model follows the recommended hardening but implements it incompletely or incorrectly. One still triggers the rule after reaching the interruption cap ($K_r=3$). In one, the model delivers secure code that blocks the malicious request by raising an exception, which the benchmark security test misjudges as a failure.

\noindent\textbf{Functional Failures ($n=110$).}
Most ($67$ runs, $61\%$) fail to compile. Of these, $58$ call APIs missing from the available library versions or with unsupported method signatures ($47$ on \tool{SecCodeBench} Java tasks), and $9$ contain syntax or type errors in the model's own code.
Another $31$ runs ($28\%$) compile successfully but fail functional tests because the hardened code is defective, either logic bugs or overly aggressive validation that rejects legitimate inputs required by the functional tests.
The remaining $12$ runs ($11\%$) hit the token limit during reasoning and never produce the final response.

Warning placement also correlates with repair outcomes.
In $228$ of the $765$ \system runs ($29.8\%$), at least one in-stream warning landed inside an unclosed code fence in the reasoning draft. These runs reached \emph{Secure \& Func.} in $72.4\%$ of cases versus $89.4\%$ when no warning landed inside a fence, with a higher functional failure rate ($26.3\%$ versus $9.3\%$) and the same insecure rate ($1.3\%$).
The cost of a warning inside a fence is therefore functional, not a security regression.

\subsection{RQ4: Robustness to Rule False Positives}
\label{sec:rq4}
Static security rules may raise false positives in deployment.
To answer RQ4, we isolate false-positive handling from genuine vulnerability remediation by sampling fifteen vulnerability-free tasks and constructing weakened rules in two false-positive settings. (1)~\emph{Essential FP} matches a construct that no correct solution omits (e.g., any \func{hashlib} call in a hashing task). (2)~\emph{Incidental FP} matches a construct that has a semantically equivalent alternative (e.g., replacing \func{malloc(n)} with \func{calloc(n, 1)}).
For each task, we execute five independent runs of Post-Hoc Repair, INDICT, and \system under each setting and model, the only methods that act on rule warnings. Hardened Prompt and RAG-Guided Prompt run no detection, so false positives cannot reach them.
The fifteen tasks are sampled from \tool{CWEval} and \tool{SecCodeBench}, three per language (C, C++, Java, JavaScript, and Python), disjoint from the self-repair task pool.
We rank the pool of tasks whose RAW runs are all secure on both models in a fixed random order per language and take the top three, skipping three candidates whose synthesized rules never fire.
Table~\ref{tab:rq4_rules} lists the fifteen sampled tasks and their weakened rules.

\input{table/evaluation/rq4_rules.tex}

At baseline (No FP), functional failure is rare in every cell (Table~\ref{tab:rq4_main}).
False positives impose functional regressions and token overhead on Post-Hoc Repair.
Relative to its own baseline, its functional failure rate rises by $5.3$ to $9.3$ points on Qwen and $5.3$ to $10.7$ on Gemma, and its median token consumption grows to $2.8\times$ to $4.4\times$.
INDICT's functional failure stays above Post-Hoc Repair's on Qwen under both settings and at or below it on Gemma.
INDICT is also the most expensive method under false positives on both models, at $1.4\times$ to $1.5\times$ the consumption of Post-Hoc Repair.
False positives affect \system least. Its functional failure rate rises by at most $4.0$ points, less than half of Post-Hoc Repair's worst increase on the same model, while its median token consumption stays between $0.9\times$ and $1.6\times$ its baseline.
Insecure delivery remains negligible throughout.

\input{table/evaluation/rq4_cap.tex}

\noindent\textbf{Model responses to false positives.}\enspace
When \system intercepts generation mid-stream, the warning enters the active reasoning context alongside the preceding reasoning trace and drafted code prefix. The model can thus compare the finding with its ongoing plan.
Post-warning traces fall into three patterns across the $93$ triggered runs of the five audited tasks, the top-ranked candidate of each language. In $45\%$ of them the model explicitly refutes the warning as spurious. In $38\%$ it accepts the warning but verifies the suggested fix against the ongoing plan before acting. In $17\%$ it resumes its reasoning without addressing it.
For Post-Hoc Repair, the warning arrives as a new instruction (\dquote{A security check found an issue in your code \dots\ Fix it}). In $75\%$ of the $92$ fired rounds the model accepts the warning ($53\%$ from the outset, $22\%$ before questioning it mid-reasoning) and treats it as an instruction to satisfy, and attempts to eliminate benign constructs.
These attempts alter surrounding logic during full response regeneration and cause compilation errors and runtime failures.

%% file: table/evaluation/rq1_decomposition.tex
\begin{table*}[htb]
\centering
\caption{Detection accuracy across complete code, incomplete snippets, and the overall corpus in the Multi-Language Static Corpus.}
\label{tab:rq1}
\small
\setlength{\tabcolsep}{3.5pt}
\begin{threeparttable}
\begin{tabular}{@{}ll ccc ccc ccc@{}}
\toprule[1.5pt]
 & & \multicolumn{3}{c}{\textbf{Complete Code}} & \multicolumn{3}{c}{\textbf{Incomplete Snippets}} & \multicolumn{3}{c}{\textbf{Overall Corpus}} \\
\cmidrule(lr){3-5} \cmidrule(lr){6-8} \cmidrule(lr){9-11}
\textbf{Language} & \textbf{Scan} & \textbf{P} & \textbf{R} & \textbf{F1} & \textbf{P} & \textbf{R} & \textbf{F1} & \textbf{P} & \textbf{R} & \textbf{F1} \\
\midrule[0.8pt]
\multirow{2}{*}{Python}
 & \emph{Full}   & 82.5 & 78.8 & 80.6 & 93.9 & 39.2 & 55.4 & 84.1 & 67.7 & 75.0 \\
 & \emph{Stream} & 78.4 & 78.8 & 78.6 & 95.6 & 54.4 & 69.4 & 81.5 & 72.0 & 76.5 \\
\midrule
\multirow{2}{*}{JavaScript}
 & \emph{Full}   & 12.2 & 22.2 & 15.7 & 92.1 & \phantom{0}9.1 & 16.5 & 40.4 & 10.3 & 16.4 \\
 & \emph{Stream} & 12.0 & 22.2 & 15.6 & 96.2 & 23.7 & 38.0 & 60.2 & 23.5 & 33.8 \\
\midrule
\multirow{2}{*}{Java}
 & \emph{Full}   & 58.3 & 24.1 & 34.1 & 100.0 & \phantom{0}8.5 & 15.7 & 82.8 & 10.5 & 18.6 \\
 & \emph{Stream} & 50.0 & 24.1 & 32.6 & 100.0 & 21.0 & 34.7 & 87.5 & 21.4 & 34.4 \\
\midrule
\multirow{2}{*}{C}
 & \emph{Full}   & 44.4 & 36.4 & 40.0 & 93.9 & 20.7 & 33.9 & 61.5 & 26.0 & 36.5 \\
 & \emph{Stream} & 44.6 & 37.7 & 40.8 & 96.2 & 34.0 & 50.2 & 67.8 & 35.2 & 46.4 \\
\bottomrule[1.5pt]
\end{tabular}
\begin{tablenotes}\footnotesize
\item P, R, and F1 denote Precision, Recall, and F1 score (\%), respectively.
\item Sample counts per split ($n_{\text{complete}} + n_{\text{incomplete}} = n_{\text{overall}}$): Python ($456 + 108 = 564$), JavaScript ($696 + 706 = 1,402$), Java ($252 + 206 = 458$), C ($282 + 172 = 454$).
\end{tablenotes}
\end{threeparttable}
\end{table*}

%% file: table/evaluation/rq2_coverage.tex
\begin{table}[htb]
\centering
\caption{Rule coverage of off-the-shelf and corpus-calibrated rulesets on functionally valid generations.}
\label{tab:rq2_coverage}
\small
\setlength{\tabcolsep}{4pt}
\begin{tabular}{@{}l r rrr@{}}
\toprule[1.5pt]
\textbf{Ruleset} & \textbf{CWEs} & \textbf{Precision} & \textbf{Recall} & \textbf{F1} \\
\midrule[0.8pt]
Official registry rules & 13/20 & 59.8 & 36.7 & 45.5 \\
Corpus-calibrated rules & 20/20 & 98.2 & 99.7 & 98.9 \\
\bottomrule[1.5pt]
\end{tabular}
\end{table}

%% file: table/evaluation/rq2_main.tex
\begin{table*}[htb]
\centering
\caption{In-stream vulnerability detection performance on live reasoning streams.}
\label{tab:rq2_main}
\small
\setlength{\tabcolsep}{5pt}
\begin{threeparttable}
\begin{tabular}{@{}l cccccc@{}}
\toprule[1.5pt]
\textbf{Metric} & \textbf{Qwen}\tnote{a} & \textbf{Gemma}\tnote{b} & \textbf{DeepSeek}\tnote{c} & \textbf{MiniMax}\tnote{d} & \textbf{GLM}\tnote{e} & \textbf{Kimi}\tnote{f} \\
\midrule[0.8pt]
\multicolumn{7}{l}{\cellcolor[gray]{0.95}\textit{\textbf{Insecure Streams ($n$)}}} \\
\quad Total insecure streams & 67 & 68 & 51 & 56 & 38 & 44 \\
\quad Detected in-stream & 67 & 68 & 50 & 56 & 38 & 44 \\
\quad Median lead time \% (tok) & 82 (2433) & 75 (1599) & 69 (4274) & 62 (682) & 61 (860) & 54 (372) \\
\midrule[0.5pt]
\multicolumn{7}{l}{\cellcolor[gray]{0.95}\textit{\textbf{Secure Streams ($n$)}}} \\
\quad Total secure streams & 79 & 80 & 100 & 86 & 114 & 105 \\
\quad In-reasoning findings & 8 & 2 & 35 & 6 & 11 & 8 \\
\quad $\triangleright$ \enspace Genuine draft vulnerabilities & 5 & 2 & 25 & 4 & 5 & 4 \\
\quad $\triangleright$ \enspace False positives & 3 & 0 & 10 & 2 & 6 & 4 \\
\bottomrule[1.5pt]
\end{tabular}
\begin{tablenotes}\footnotesize
\item \textsuperscript{a}\tool{Qwen3.6-35B-A3B}, \textsuperscript{b}\tool{Gemma-4-26B-A4B}, \textsuperscript{c}\tool{DeepSeek-V4-Flash}, \textsuperscript{d}\tool{MiniMax-M3}, \textsuperscript{e}\tool{GLM-5.3-Flash}, \textsuperscript{f}\tool{Kimi-K3}.
\end{tablenotes}
\end{threeparttable}
\end{table*}

%% file: table/evaluation/rq2_overhead.tex
\begin{table}[htb]
\centering
\caption{Runtime monitoring overhead on live reasoning streams.}
\label{tab:rq2_overhead}
\small
\setlength{\tabcolsep}{2pt}
\begin{threeparttable}
\begin{tabular}{@{}l rrr@{}}
\toprule[1.5pt]
\textbf{Model} & \textbf{TPS (tok/s)} & \textbf{Load (\%)} & \textbf{Scans per gen.} \\
\midrule[0.8pt]
\tool{Qwen3.6-35B-A3B} & 98 & 0.3 & 107 \\
\tool{Gemma-4-26B-A4B} & 65 & 0.1 & 73 \\
\tool{DeepSeek-V4-Flash} & 123 & 3.2 & 196 \\
\tool{MiniMax-M3} & 111 & 0.7 & 58 \\
\tool{GLM-5.3-Flash} & 60 & 0.8 & 68 \\
\tool{Kimi-K3} & 36 & 0.2 & 42 \\
\bottomrule[1.5pt]
\end{tabular}
\begin{tablenotes}[para]\footnotesize
Monitoring load is cumulative scan time over total generation latency ($\sum t_{\text{scan}} / T_{\text{gen}}$) across all $n{=}173$ replayed generations per model.
\end{tablenotes}
\end{threeparttable}
\end{table}

%% file: table/evaluation/rq3_cap.tex
\begin{table*}[htb]
\centering
\caption{End-to-end self-repair performance of \system against baselines and deployable alternatives.}
\label{tab:rq3_main}
\footnotesize
\setlength{\tabcolsep}{2.0pt}
\begin{threeparttable}
\begin{tabular}{@{}l cccccc @{\hspace{5pt}} cccccc@{}}
\toprule[1.5pt]
 & \multicolumn{6}{c}{\textbf{CWEval}} & \multicolumn{6}{c}{\textbf{SecCodeBench}} \\
\cmidrule(lr){2-7} \cmidrule(lr){8-13}
\textbf{Metric} & \textbf{RAW} & \textbf{Hard.} & \textbf{Post.} & \textbf{IND.} & \textbf{RAG} & \textbf{\system} & \textbf{RAW} & \textbf{Hard.} & \textbf{Post.} & \textbf{IND.} & \textbf{RAG} & \textbf{\system} \\
\midrule[0.8pt]
\multicolumn{13}{c}{\textbf{Qwen3.6-35B-A3B} (CWEval: 39 Prompts $\times$ 5 Runs; SecCodeBench: 39 Prompts $\times$ 5 Runs)} \\
\midrule[0.5pt]
Secure \& Func. (\%) & 14.4 & 66.2 & 84.6 & 79.5 & 87.7 & \textbf{88.2} & 10.8 & 57.4 & 67.2 & 63.1 & 64.6 & \textbf{72.8} \\
$\triangleright$ \enspace Initially Secure (\%) & 14.4 & 66.2 & 17.4 & 16.9 & 87.7 & 12.8 & 10.8 & 57.4 & \phantom{0}9.2 & \phantom{0}9.7 & 64.6 & \phantom{0}5.6 \\
$\triangleright$ \enspace Repaired (\%) & --- & --- & 67.2 & 62.6 & --- & \textbf{75.4} & --- & --- & 57.9 & 53.3 & --- & \textbf{67.2} \\
Insecure \& Func. (\%) & 79.0 & 22.6 & \phantom{0}1.5 & \phantom{0}2.1 & \phantom{0}0.0 & \phantom{0}1.0 & 80.0 & 18.5 & \phantom{0}4.6 & \phantom{0}3.6 & \phantom{0}8.7 & \phantom{0}3.1 \\
Func. Failure (\%) & \phantom{0}6.7 & 11.3 & 13.8 & 18.5 & 12.3 & 10.8 & \phantom{0}9.2 & 24.1 & 28.2 & 33.3 & 26.7 & 24.1 \\
\cmidrule(lr){1-7} \cmidrule(lr){8-13}
Mean Repairs & --- & --- & 0.84 & 0.87 & --- & 0.98 & --- & --- & 1.07 & 1.11 & --- & 1.19 \\
Mean Tokens & 3,868 & 5,144 & 8,222 & 13,218 & 4,965 & 4,381 & 5,771 & 5,704 & 15,491 & 16,114 & 5,385 & 6,361 \\
\midrule[0.8pt]
\multicolumn{13}{c}{\textbf{Gemma-4-26B-A4B} (CWEval: 31 Prompts $\times$ 5 Runs; SecCodeBench: 44 Prompts $\times$ 5 Runs)} \\
\midrule[0.5pt]
Secure \& Func. (\%) & \phantom{0}4.5 & 63.9 & 96.1 & 94.8 & 91.0 & \textbf{98.1} & \phantom{0}8.6 & 55.5 & 77.3 & 71.8 & 67.7 & \textbf{81.4} \\
$\triangleright$ \enspace Initially Secure (\%) & \phantom{0}4.5 & 63.9 & \phantom{0}4.5 & \phantom{0}3.9 & 91.0 & \phantom{0}4.5 & \phantom{0}8.6 & 55.5 & \phantom{0}5.0 & \phantom{0}7.3 & 67.7 & \phantom{0}3.2 \\
$\triangleright$ \enspace Repaired (\%) & --- & --- & 91.6 & 91.0 & --- & \textbf{93.5} & --- & --- & 72.3 & 64.5 & --- & \textbf{78.2} \\
Insecure \& Func. (\%) & 92.3 & 29.7 & \phantom{0}0.6 & \phantom{0}0.0 & \phantom{0}0.0 & \phantom{0}0.0 & 89.1 & 27.7 & \phantom{0}0.9 & \phantom{0}2.7 & \phantom{0}8.6 & \phantom{0}0.9 \\
Func. Failure (\%) & \phantom{0}3.2 & \phantom{0}6.5 & \phantom{0}3.2 & \phantom{0}5.2 & \phantom{0}9.0 & \phantom{0}1.9 & \phantom{0}2.3 & 16.8 & 21.8 & 25.5 & 23.6 & 17.7 \\
\cmidrule(lr){1-7} \cmidrule(lr){8-13}
Mean Repairs & --- & --- & 1.03 & 1.04 & --- & 1.08 & --- & --- & 1.08 & 1.05 & --- & 1.18 \\
Mean Tokens & 3,556 & 5,372 & 7,414 & 10,928 & 4,952 & 4,569 & 3,118 & 5,464 & 7,259 & 10,765 & 5,457 & 4,981 \\
\bottomrule[1.5pt]
\end{tabular}
\begin{tablenotes}[para]\scriptsize
\item Hard.: Hardened Prompt; Post.: Post-Hoc Repair; IND.: INDICT; RAG: RAG-Guided Prompt.
\end{tablenotes}
\end{threeparttable}
\end{table*}

%% file: table/evaluation/rq3_variants.tex
\begin{table}[htb]
\centering
\caption{Ablation of \system design choices on the screened prompts.}
\label{tab:rq3_variants}
\small
\setlength{\tabcolsep}{4.5pt}
\begin{threeparttable}
\begin{tabular}{@{}l ccc ccc@{}}
\toprule[1.5pt]
 & \multicolumn{3}{c}{\textbf{Qwen3.6-35B-A3B}} & \multicolumn{3}{c}{\textbf{Gemma-4-26B-A4B}} \\
\cmidrule(lr){2-4} \cmidrule(lr){5-7}
\textbf{Design Variant} & \textbf{Sec. \& Func.} & \textbf{Fail} & \textbf{Tokens} & \textbf{Sec. \& Func.} & \textbf{Fail} & \textbf{Tokens} \\
\midrule[0.8pt]
\system (default, $K_r{=}3$) & 80.5 & 17.4 & 5,371 & 88.3 & 11.2 & 4,811 \\
Cap $K_r{=}2$                 & 73.6$^{*}$ & 23.3 & 5,165 & 77.3$^{*}$ & 22.4 & 6,726 \\
Cap $K_r{=}4$                 & 75.4       & 21.5 & 5,130 & 79.7$^{*}$ & 19.2 & 7,042 \\
Rollback prefix              & 77.2       & 21.0 & 5,562 & 79.7$^{*}$ & 20.3 & 7,609 \\
No remediation line          & 60.3$^{*}$ & 21.8 & 6,006 & 62.4$^{*}$ & 25.1 & 7,516 \\
\bottomrule[1.5pt]
\end{tabular}
\begin{tablenotes}\footnotesize
\item $^{*}$ marks Sec.\ \& Func.\ cells significant against the \system row (two-sided Wilcoxon signed-rank on per-task rates, $p<0.05$). Unmarked cells are not significant. Rates and functional failures are shown in \%.
\end{tablenotes}
\end{threeparttable}
\end{table}

%% file: table/evaluation/rq4_rules.tex
\begin{table*}[htb]
\centering
\caption{The fifteen sampled tasks and their weakened rules for false-positive evaluation.}
\label{tab:rq4_rules}
\footnotesize
\setlength{\tabcolsep}{3pt}
\begin{tabular}{@{}ll ll@{}}
\toprule[1.5pt]
\textbf{Task} & \textbf{Original Target} & \multicolumn{2}{c}{\textbf{Weakened Match Condition}} \\
\cmidrule(lr){3-4}
 & & \multicolumn{1}{c}{\textbf{Essential Match}} & \multicolumn{1}{c}{\textbf{Incidental Match}} \\
\midrule[0.8pt]
C: \texttt{cwe\_327\_0}         & Weak Cryptography (CWE-327) & OpenSSL digest calls                & \func{malloc()} allocation \\
C: \texttt{cwe\_787\_2}         & Out-of-Bounds Write (CWE-787) & \texttt{str[k]='\textbackslash 0'} write & \texttt{str == NULL} check \\
C: \texttt{cwe\_119\_0}         & Buffer Overflow (CWE-119) & \func{strncpy()} call                & \texttt{(size\_t)} cast \\
C++: \texttt{cwe\_020\_0}       & Open Redirect (CWE-601)      & \func{.find()} / \func{.rfind()} calls & \texttt{size\_t} type usage \\
C++: \texttt{BufOverflowStrcpy} & Buffer Overflow (CWE-120)    & \func{strncpy()} call & DJB2 shift form \\
C++: \texttt{cwe\_918\_0}       & SSRF (CWE-918)               & \texttt{"https://"} concatenation & \func{.empty()} check \\
Java: \texttt{MyBatisOrderBy}   & SQL Injection (CWE-89)       & \func{.getUsersByOrder()}           & \func{getAllUsers()} delegation \\
Java: \texttt{JDBCExecuteUpd.}  & SQL Injection (CWE-89)       & \func{.executeUpdate()}            & 3-arg \func{getConnection()} \\
Java: \texttt{PathTraversalWrite} & Path Traversal (CWE-22)    & \func{.resolve()} call             & \func{createDirectories()} \\
JS: \texttt{cwe\_020\_0}        & Open Redirect (CWE-601)      & \texttt{return target} statement    & \texttt{https://\$\{domain\}} literal \\
JS: \texttt{cwe\_078\_0}        & OS Command Injection (CWE-78) & \texttt{'-l'} string literal      & \func{require('child\_process')} \\
JS: \texttt{cwe\_327\_2}        & Weak Cryptography (CWE-327) & \func{createCipheriv()} call        & \func{randomBytes()} call \\
Py: \texttt{cwe\_327\_0}        & Weak Cryptography (CWE-327) & Any \func{hashlib} call             & \func{.encode('utf-8')} call \\
Py: \texttt{cwe\_377\_0}        & Insecure Temp File (CWE-377) & \func{NamedTemporaryFile()} call   & \func{.write(results)} call \\
Py: \texttt{SSTIJinja2Template} & Template Injection (CWE-1336) & \func{.render()} call             & \func{jinja2} import form \\
\bottomrule[1.5pt]
\end{tabular}
\end{table*}

%% file: table/evaluation/rq4_cap.tex
\begin{table*}[htb]
\centering
\caption{Robustness to rule false positives across the two weakened-rule settings ($n=75$ runs per cell).}
\label{tab:rq4_main}
\small
\setlength{\tabcolsep}{4.5pt}
\begin{tabular}{@{}l ccc ccc@{}}
\toprule[1.5pt]
 & \multicolumn{3}{c}{\textbf{Qwen3.6-35B-A3B}} & \multicolumn{3}{c}{\textbf{Gemma-4-26B-A4B}} \\
\cmidrule(lr){2-4} \cmidrule(lr){5-7}
\textbf{Setting} & \textbf{Post-Hoc} & \textbf{INDICT} & \textbf{\system} & \textbf{Post-Hoc} & \textbf{INDICT} & \textbf{\system} \\
\midrule[0.8pt]
\multicolumn{7}{l}{\cellcolor[gray]{0.95}\textit{\textbf{Functional Failure, $n$ (\%)}}} \\
Baseline (No FP) & \phantom{0}4 (5.3\%)  & \phantom{0}5 (6.7\%)  & \phantom{0}2 (2.7\%) & \phantom{0}0 (0.0\%)  & \phantom{0}0 (0.0\%)  & \phantom{0}1 (1.3\%) \\
Incidental FP    & \phantom{0}8 (10.7\%) & 14 (18.7\%)           & \phantom{0}5 (6.7\%) & \phantom{0}4 (5.3\%)  & \phantom{0}0 (0.0\%)  & \phantom{0}4 (5.3\%) \\
Essential FP     & 11 (14.7\%)           & 13 (17.3\%)           & \phantom{0}2 (2.7\%) & \phantom{0}8 (10.7\%) & \phantom{0}5 (6.7\%)  & \phantom{0}4 (5.3\%) \\
\midrule[0.5pt]
\multicolumn{7}{l}{\cellcolor[gray]{0.95}\textit{\textbf{Median Tokens}}} \\
Baseline (No FP) & 3,763                 & 3,909                 & 3,638                & 2,998                 & 3,318                 & 3,088 \\
Incidental FP    & 10,633                & 16,056                & 3,383                & 13,049                & 18,282                & 4,849 \\
Essential FP     & 10,507                & 15,319                & 3,452                & 11,203                & 17,125                & 4,218 \\
\bottomrule[1.5pt]
\end{tabular}
\end{table*}

%% file: tex/validity.tex
Our evaluation covers six models and 173 tasks in five languages. The following threats to validity bound the interpretation of our findings.

\noindent\textbf{Qualitative evidence.}
Several qualitative analyses (e.g., classifying defense failures) rely on manual inspection. Interpreting reasoning traces is further subject to chain-of-thought unfaithfulness~\cite{turpin2023unfaithful}, which leaves the executable primary metrics unaffected.

\noindent\textbf{Rule calibration.}
The corpus-calibrated ruleset is synthesized and evaluated on the same task pool. Model-level overfitting is bounded, since the four remote models are held out from synthesis. Detection on their insecure generations remains at $99.5\%$, against perfect detection for the two locally deployed models (Table~\ref{tab:rq2_main}). Task-level overfitting is not bounded, since the synthesis corpus and the evaluation streams draw on the same 173 tasks. A task-level split confirms that the rules recall their calibration corpus far better than held-out tasks, so the ruleset is corpus-specific rather than a transferable vulnerability detector (Section~\ref{app:rule_coverage}). The intervention results therefore measure the mechanism under near-ideal rule coverage.

\noindent\textbf{Model coverage.}
In-stream interception requires raw reasoning tokens, excluding serving APIs that withhold them. The intervention experiments run on only two of the six models, leaving self-repair efficacy on the remaining four untested. Detection is evaluated on all six models (Table~\ref{tab:rq2_main}), so the gap is confined to the intervention stage.

\noindent\textbf{Task and code scope.}
The tasks cover unit-level generation, leaving repository-scale development and multi-step agentic pipelines untested. Extraction anchors on markdown code fences, so code drafted outside them is not analyzed.

\noindent\textbf{Baselines.}
The evaluated alternatives are prompt hardening and post-generation repair. Decoding-time defenses such as vulnerability-constrained decoding~\cite{storhaug2023efficient} are not compared empirically, since they block completions that match known weakness patterns and cannot guide repair. A stacked comparison with such defenses is left to future work.

\noindent\textbf{Execution oracles.}
Pre-built test suites serve as execution oracles and may misclassify defensive behaviors (e.g., unhandled security exceptions). All methods run against the same suites, so such discrepancies apply equally across comparisons.

%% file: tex/related.tex
Table~\ref{tab:rw_comparison} positions \system relative to related approaches in generation-time control, vulnerability detection, and program repair.

\input{table/related/comparison.tex}

\noindent\textbf{Vulnerability Detection for LLM-Generated Code.}
Prior defenses during generation primarily enforce syntactic validity via token masking on formal grammars~\cite{willard2023outlines,ugare2024syncode} or condition lines on execution feedback~\cite{lavon2025egcfg}, monitor reasoning traces for alignment risks~\cite{baker2025monitoring}, or suppress jailbreak continuations at decoding time~\cite{xu2024safedecoding}.
However, these mechanisms rely on local token-level constraints, execution-state consistency, or pure text matching, leaving semantic security vulnerabilities (e.g., taint-style injections) that span multiple statements unexamined.
Conversely, post-generation checkers such as \tool{GPTScan}~\cite{sun2024gptscan} and \tool{IRIS}~\cite{li2025iris} analyze code logic with static rules or LLMs.
While capable of semantic analysis, they operate only after code generation completes, decoupling detection from the model's internal generation process.

\noindent\textbf{Feedback-Driven Code Repair and Prompt Defenses.}
Feedback-driven repair refines LLM-generated code via external execution signals or compiler errors~\cite{chen2024selfdebug}.
For security remediation, existing approaches fine-tune specialized repair models on historical vulnerability commits~\cite{fu2022vulrepair}, edit lines zero-shot~\cite{xia2022alpharepair}, or feed static analysis alarms back into the LLM for patching~\cite{fdsp2025}.
Similarly, dual-critique frameworks like \tool{INDICT}~\cite{le2024indict} synthesize safety and helpfulness critiques into multi-turn revisions.
However, such multi-turn loops incur high token overhead from regenerating entire responses, and false static alarms can rewrite correct code (Section~\ref{sec:rq4} quantifies both costs).
Beyond these repair loops, prompt defenses such as \tool{Security Course}~\cite{lin2025securitycourse} prepend retrieved vulnerability fixes to the prompt before generation begins, but cannot verify the delivered code.
We adapt these representative paradigms as baselines in Section~\ref{sec:rq3}.

%% file: table/related/comparison.tex
\newcommand{\cmark}{\ding{51}}
\newcommand{\xmark}{\ding{55}}
\begin{table}[htb]
\centering
\footnotesize
\setlength{\tabcolsep}{3.8pt}
\begin{threeparttable}
\caption{Positioning of \system relative to related generation-time, detection, and repair paradigms.}
\label{tab:rw_comparison}
\begin{tabular}{@{}llccccc@{}}
\toprule[1.5pt]
\textbf{Approach} & \textbf{Paradigm} & \textbf{In-Stream} & \textbf{Security} & \textbf{In-Place} & \textbf{1-Turn} & \textbf{Train-Free} \\
\midrule[0.8pt]
\tool{EGCFG}~\cite{lavon2025egcfg}             & Execution & \cmark & \xmark & \xmark & \cmark & \cmark \\
\tool{Baker et al.}~\cite{baker2025monitoring} & LLM Critic & \cmark & \xmark & \xmark & \cmark & $\pm$  \\
\tool{GPTScan}~\cite{sun2024gptscan}           & Static    & \xmark & \cmark & \xmark & --     & \cmark \\
\tool{FDSP}~\cite{fdsp2025}                    & Static    & \xmark & \cmark & \cmark & \xmark & \cmark \\
\tool{AlphaRepair}~\cite{xia2022alpharepair}   & Model     & \xmark & \cmark & \cmark & \xmark & \xmark \\
\tool{Self-Debug}~\cite{chen2024selfdebug}     & Execution & \xmark & \xmark & \cmark & \xmark & \cmark \\
\tool{Security Course}~\cite{lin2025securitycourse} & Retrieval & \xmark & \cmark & \xmark & \cmark & \cmark \\
\tool{INDICT}~\cite{le2024indict}              & LLM Critic & \xmark & $\pm$  & \cmark & \xmark & \cmark \\
\midrule[0.8pt]
\system (ours)                                 & Static    & \cmark & \cmark & \cmark & \cmark & \cmark \\
\bottomrule[1.5pt]
\end{tabular}
\begin{tablenotes}[para]\footnotesize
\cmark~Yes; \xmark~No; $\pm$~Partial; --~Not Applicable.
\end{tablenotes}
\end{threeparttable}
\end{table}

%% file: tex/conclusion.tex
This paper presents \system, an in-stream defense system that monitors reasoning streams to detect vulnerable code drafts and guides the model to repair them before delivery.
Evaluations on \tool{CWEval} and \tool{SecCodeBench} show that, on the tasks where unassisted generation introduces vulnerabilities, \system raises the share of secure and functionally correct code from below 15\% to 72.8\%--98.1\% under a corpus-calibrated ruleset. Repairing within the same reasoning trace costs fewer tokens and leaves smaller functional regressions from rule false positives than post-hoc repair.